\documentclass{nature}
\usepackage{graphicx}
\usepackage{aasymbols}
\usepackage{hyperref}
\usepackage[switch, modulo]{lineno}
\modulolinenumbers[1]

\usepackage{xpatch}

\usepackage[dvipsnames]{xcolor}

\newcommand{\Line}[3]{\Ion{#1}{#2}~#3}

\newcommand{\Ion}[2]{#1{\,\textsc{\small#2}}}

\title{Accretion of a giant planet onto a white dwarf}

\author{
Boris T. G\"ansicke$^{1,2}$, 
Matthias R. Schreiber$^{3}$,
Odette Toloza$^{1}$,
Nicola P. Gentile Fusillo$^{1}$, \\
Detlev Koester$^{4}$ \&
Christopher J. Manser$^{1}$}

\begin{document}

\maketitle

\begin{affiliations}
\item Department of Physics, University of Warwick, Coventry CV4 7AL, UK
\item Centre for Exoplanets and Habitability, University of Warwick, Coventry CV4 7AL, UK
\item Institute of Physics and Astronomy, Millennium Nucleus for Planet Formation (NPF), Universidad de Valpara\'iso, Av. Gran Bretana 1111, Valpara\'iso, Chile
\item Institut f\"ur Theoretische Physik und Astrophysik, Universit\"at Kiel, 24098 Kiel, Germany
\end{affiliations}

\begin{abstract}
The detection of a dust disc around G29-38\cite{zuckerman+becklin87-1}
and transits from debris orbiting
WD\,1145+017\cite{vanderburgetal15-1} confirmed that the photospheric
trace metals found in many white dwarfs\cite{koesteretal14-1} arise
from the accretion of tidally disrupted
planetesimals\cite{jura03-1}. The  composition of these planetesimals
is similar to that of rocky bodies in the inner solar
system\cite{zuckermanetal07-1}. Gravitationally scattering
planetesimals towards the white dwarf requires the presence of more
massive bodies\cite{frewen+hansen14-1}, yet no planet has so far been
detected at a white dwarf. Here we report optical spectroscopy of a
$\simeq27\,750$\,K hot white dwarf that is accreting from a
circumstellar gaseous disc composed of hydrogen, oxygen, and sulphur
at a rate of $\simeq3.3\times10^9\,\mathrm{g\,s^{-1}}$. The
composition of this disc is unlike all other known planetary debris
around white dwarfs\cite{xuetal17-1}, but resembles predictions for
the makeup of deeper atmospheric layers of icy giant planets, with
H$_2$O and H$_2$S being major constituents. A giant planet orbiting a
hot white dwarf with a semi-major axis of $\simeq15$ solar radii will
undergo significant evaporation with expected mass loss rates
comparable to the accretion rate onto the white dwarf. The orbit of
the planet is most likely the result of gravitational interactions,
indicating the presence of additional planets in the system. We infer
an occurrence rate of spectroscopically detectable giant planets in close orbits around white dwarfs of $\simeq10^{-4}$.
\end{abstract}

\noindent
WD\,J091405.30+191412.25 (WD\,J0914+1914) was initially classified as a close interacting white dwarf binary on the basis of a weak H$\alpha$ emission line detected in its spectrum obtained by the Sloan Digital Sky Survey (SDSS)\cite{gentile-fusilloetal15-1}. Upon closer inspection of this spectrum, we identified additional emission lines of oxygen (\Ion{O}{i} at wavelengths 7,774\,\AA\ and 8,446\,\AA), and an emission line near 4,068\,\AA\ that we tentatively identified as [\Ion{S}{ii}]. The line flux ratios of the hydrogen and oxygen lines are extremely atypical for any white dwarf binary, casting doubt on the published classification.  

We obtained deep spectroscopy of this star using the X-Shooter spectrograph on the Very Large Telescope of the European Southern Observatory (see Fig.~1) which confirms the presence of [\Ion{S}{ii}] (4,068\,\AA), and contains additional emission lines of [\Ion{O}{i}] (6,300\,\AA\ and 6,363\,\AA) as well as a blend of \Ion{O}{i} and \Ion{S}{i} lines near 9,200\,\AA. 

The double-peaked morphology of the H$\alpha$ and the \Ion{O}{i} (8,446\,\AA) emission lines (see Fig.~1) indicates an origin in a circumstellar gas disc\cite{horne+marsh86-1}, reminiscent of several white dwarfs with dusty and gaseous planetary debris discs\cite{gaensickeetal06-3, melisetal12-1}. However, the spectra of all known gaseous debris discs are dominated by the emission lines of the \Ion{Ca}{ii} triplet (8,600\,\AA), with weaker lines of \Ion{Fe}{ii}, which are absent in the X-Shooter observations of WD\,J0914+1914. Moreover, none of the other gaseous debris discs around white dwarfs show H$\alpha$ emission. 

The X-shooter spectrum of WD\,J0914+1914 displays strong Balmer lines, implying a hydrogen-dominated atmosphere, as well as numerous sharp absorption lines of oxygen and sulphur (see Fig.\,2). We determined the white dwarf's effective temperature of $T_\mathrm{eff}=27,743\pm310$\,K and a surface gravity of $\log g=7.85\pm0.06$ from the well flux-calibrated SDSS spectra (see Extended Data Fig.\,1 and Extended Data Table\,1). Fixing these two atmospheric parameters, we measured the photospheric abundances of oxygen and sulphur, $\log(\mathrm{O/H)}=-3.25\pm0.20$ and $\log\mathrm{(S/O)}=-4.15\pm0.20$, and derived upper limits for twelve additional elements (see Fig.\,3 and Extended Data Table\,2).  WD\,J0914+1914 is accreting at a rate of $\simeq3.3\times10^9\,\mathrm{g\,s^{-1}}$, which is among the highest of all hydrogen-atmosphere white dwarfs polluted by planetary debris\cite{koesteretal14-1}. However, the measured accretion rate in  WD\,J0914+1914 includes only oxygen and sulphur, and the influxes of these two elements are an order of magnitude larger than in any other of these systems.  If thermohaline mixing or convective overshoot are efficient in the atmosphere of WD\,J0914+1914, the accretion rate could be an order of magnitude higher\cite{bauer+bildsten19-1}.

We used the spectral synthesis code \textsc{Cloudy}\cite{ferlandetal17-1} to model the photoionisation of the circumstellar gas disc by the intense ultraviolet flux from the white dwarf (see Methods and Extended Data Fig.\,2 \& 3). The emission lines, which are Doppler-broadened by the Keplerian rotation in the disc\cite{horne+marsh86-1}, originate from a gaseous disc extending $\simeq1-10\,\mathrm{R}_\odot$ (where $\mathrm{R}_\odot$ is the radius of the Sun, see Extended Data Fig.\,2 \& 4) from the white dwarf, at a density of $\rho\simeq10^{-11.3}\,\mathrm{g\,cm^{-3}}$. The relative abundances of oxygen and sulphur derived from this model, $\log(\mathrm{S/O})=-0.5$ are consistent with those measured from the photospheric analysis. Hydrogen in the disc is strongly depleted with respect to oxygen and sulphur, $\log(\mathrm{O/H})=0.29$ and $\log(\mathrm{S/H})=-0.21$. The non-detection of emission lines from other elements apart from hydrogen, oxygen and sulphur allows stringent upper limits to be placed on the abundances of sodium, silicon, calcium, and iron in the disc (see Fig.\,3 and Extended Data Table\,2). 

The abundances of the gaseous circumstellar disc, and of the trace metals in the photosphere of WD\,J0914+1914 are distinctly non-solar and inconsistent with accretion from the wind of a low-mass companion star\cite{pyrzasetal12-1}. A stellar companion is also ruled out by the stringent upper limits on the radial velocity variations of the white dwarf and the absence of an infrared excess (see Methods). In contrast to the white dwarfs known to be contaminated by planetary debris, the material accreted by WD\,J0914+1914 is extremely depleted in the major rock-forming elements magnesium, silicon, calcium and iron with respect to the bulk Earth, and the circumstellar disc at WD\,J0914+1914 is much larger than the canonical Roche-radius for a rocky body\cite{davidsson99-1}. Both facts argue against tidally disrupted planetesimals\cite{zuckermanetal07-1, gaensickeetal12-1} as the origin of either the gaseous disc, or the photospheric trace metals that we detected. Based on the observational evidence, WD\,J0914+1914 is a white dwarf accreting from a purely gaseous circumstellar disc, and the most plausible origin of the material in that disc is an evaporating giant planet on a close-in orbit around the white dwarf. 

The abundances of WD\,J0914+1914 are reminiscent of the deeper layers of the ice giants in the solar system. Modelling the radio and microwave spectrum of Uranus required low concentrations of ammonia (NH$_3$), and large concentrations of H$_2$O\cite{depateretal89-1}. Condensation of ammonia and hydrogen sulphide (H$_2$S) into ammonium hydrosulphide (NH$_4$SH) is potentially efficient at removing ammonia from the atmosphere. However, for a solar sulphur-to-nitrogen ratio, there is insufficient sulphur to sequester all NH$_3$ into NH$_4$SH. A plausible model for the spectrum of Uranus required H$_2$O and H$_2$S concentrations enhanced by a few hundred with respect to their solar values\cite{depateretal89-1}. H$_2$S was recently detected in the atmospheres of Uranus\cite{irwinetal18-1} and Neptune\cite{irwinetal19-1}, confirming that H$_2$S ice is a major constituent of the deeper cloud layers of icy giant planets.

Significant high-energy (extreme-ultraviolet, EUV) irradiation of Neptune-mass exo-planets results in the photo-evaporation of their atmospheres. Estimated mass loss rates of warm Neptunes with semi-major axes of a few solar radii reach $10^8-10^{10}\,\mathrm{g\,s^{-1}}$ (e.g. GJ\,436b\cite{ehrenreichetal15-1} and GJ\,3470b\cite{bourrieretal18-1}) comparable to the accretion rate we derive for WD\,J0914+1914. The high-energy stellar flux required for driving the mass loss rates of the known warm Neptunes is a few per cent of the total host star luminosity, compatible with the high-energy emission of young stars\cite{tuetal15-1}. Photo-evaporation is also the most likely process causing the mass-loss of the giant planet feeding WD\,J0914+1914. With the accretion disc extending out to $\simeq10\,\mathrm{R}_\odot$, the planet is likely located at $\simeq15\,\mathrm{R}_\odot$ (see Methods). A significant fraction of the luminosity of this moderately hot ($T_\mathrm{eff}\simeq27,750$\,K) white dwarf emerges in the EUV, which results in high-energy irradiation of the planet very similar to those of mass-losing warm Neptunes orbiting main-sequence stars. The atmospheric escape rate driven by the EUV flux of WD\,J0914+1914 may be as high as $\simeq5\times10^{11}\,\mathrm{g\,s^{-1}}$ (see Extended Data Fig.\,5 and Methods), exceeding those of the warm Neptunes GJ\,436b and GJ\,3470b\cite{ehrenreichetal15-1, bourrieretal18-1}.

A fraction of the material escaping the atmosphere of the planet remains gravitationally bound to the white dwarf, forming the circumstellar disc detected in the double-peaked emission lines. From this reservoir, the material eventually accretes onto the white dwarf, resulting in photospheric oxygen and sulphur contamination. A photoionisation model for the gaseous disc implies a strong depletion of hydrogen, which is expected to be the dominant species in the planet's atmosphere, within the circumstellar disc. In addition to its large EUV luminosity, the hot white dwarf also emits copious amounts of Ly$\alpha$ photons, substantially exceeding the solar Ly$\alpha$ flux (see Extended Data Fig.\,6 and Methods). Consequently, the inflow of hydrogen is inhibited by its large cross-section in Ly$\alpha$, strongly enhancing the abundances of oxygen and sulphur in circumstellar disc and in the accreted material.

A potential analogue to the planet at WD\,J0914+1914 is HAT-P-26b, a Neptune-mass planet\cite{hartmanetal11-1} orbiting a K-star with a period of 4.26\,d. The transmission spectrum of HAT-P-26b exhibits strong H$_2$O absorption bands, with no detection of carbon-based species\cite{wakefordetal17-1}. The carbon abundance\cite{wakefordetal17-1}  in the atmosphere of HAT-P-26b, $\log(\mathrm{C/O})<-2$, is below our detection threshold ($\log(\mathrm{C/O})<-1.55$, see Methods). A detection of carbon in the photospheric spectrum of WD\,J0914+1914 will require either substantially deeper optical spectroscopy than our 200\,min-long X-Shooter observations, or far-ultraviolet spectroscopy of the strong \Line{C}{iii}{1,175\,\AA} transition. Modelling the spectrum of HAT-P-26b predicts sulphur-based cloud-forming condensates\cite{wakefordetal17-1}, however, these are not directly detected. Despite the high temperature of WD\,J0914+1914, its small radius, $0.015\,R_\odot$ implies a luminosity that is lower than that of F, G, or K-type main-sequence host stars. Hence despite the intense EUV irradiation, a planet orbiting the white dwarf WD\,J0914+1914 will be cooler than an equivalent planet around a main-sequence star. 

Gravitational interactions in multi-planet systems can perturb planets onto orbits with pericentres close to the white dwarf, where tidal effects are likely to lead to circularisation of the orbit. Common envelope evolution provides an alternative scenario to bring a planet into a close orbit around the white dwarf\cite{nelemans+tauris98-1}, though it requires rather fine-tuned initial conditions and only works for planets more massive than Jupiter (see Methods). As the white dwarf continues to cool, the mass loss rate will gradually decrease, and become undetectable in $\simeq350$\,Myr (see Extended Data Fig.\,8). By then, the giant planet will have lost $\sim0.002$ Jupiter masses (or $\sim0.04$ Neptune masses), i.e. an insignificant fraction of its total mass. 

The ubiquitous existence of planets around white dwarfs has been indirectly implied by the frequent signatures of planetesimals scattered onto orbits crossing the Roche-radii of white dwarfs, with dynamical preference for sub-Jovian mass planets \cite{frewen+hansen14-1, mustilletal18-1}. We have inspected all $\simeq7,000$ white dwarfs\cite{gentile-fusilloetal19-1} with SDSS spectroscopy, brighter than $g=19$ and hotter than $15,000$\,K for the presence of \Ion{O}{i} (7,774 and 8,446\,\AA) emission lines, but did not identify another system that resembles WD\,J0914+1914. Spectroscopic signatures of giant planets at white dwarfs are therefore rare, but follow-up observations of the $\simeq260,000$ white dwarfs identified with \textit{Gaia}\cite{gentile-fusilloetal19-1} have the potential to discover a sufficient number of such systems to enable a comparative study of their atmospheric compositions.

\clearpage

\begin{addendum}
\item Funding for the Sloan Digital Sky Survey IV has been provided by the Alfred P. Sloan Foundation, the U.S. Department of Energy Office of Science, and the Participating Institutions. The SDSS web site is www.sdss.org. Based on observations collected at the European Organisation for Astronomical Research in the Southern Hemisphere under ESO programme 0102.C-0351(A). B.T.G. and C.J.M. were supported by the UK STFC grant ST/P000495. M.R.S. acknowledges support from the Millennium Nucleus for Planet Formation (NPF) and Fondecyt (grant 1181404). O.T. was supported by a Leverhulme Trust Research Project Grant. The research leading to these results has received funding from the European Research Council under the European Union’s Horizon 2020 research and innovation programme n. 677706 (WD3D). 

\item[Author contributions] All authors contributed to the data interpretation, discussion and writing of this article. 
B.T.G. wrote the ESO proposal, carried out the observations, and modelled the emission line profiles. 
M.R.S. developed the models for the past and future evolution of the planet, and for the photo-evaporation.
O.T. developed the \textsc{CLOUDY} model for the circumstellar disc.
O.T. and D.K. carried out the photospheric analysis. 
N.P.G.F. identified WD\,J0914+1914 as unusual white dwarf and reduced the X-Shooter data.
C.J.M. searched the SDSS spectroscopic data for additional white dwarfs exhibiting oxygen or sulphur lines. 

\item[Author Information] Reprints and permissions information is available at www.nature.com/reprints. Correspondence and requests for materials should be addressed to B.T.G. (email: boris.gaensicke@warwick.ac.uk).

\item[Competing Interests] The authors declare that they have no competing interests.

\item[Data Availability] The SDSS and X-Shooter spectra analysed in this paper are available from the SDSS and ESO archives.

\item[Code Availability] \textsc{Cloudy} is publically available. The model atmosphere code of D. Koester is subject to restricted availability.

\end{addendum}


\clearpage
\centerline{\includegraphics[width=120mm]{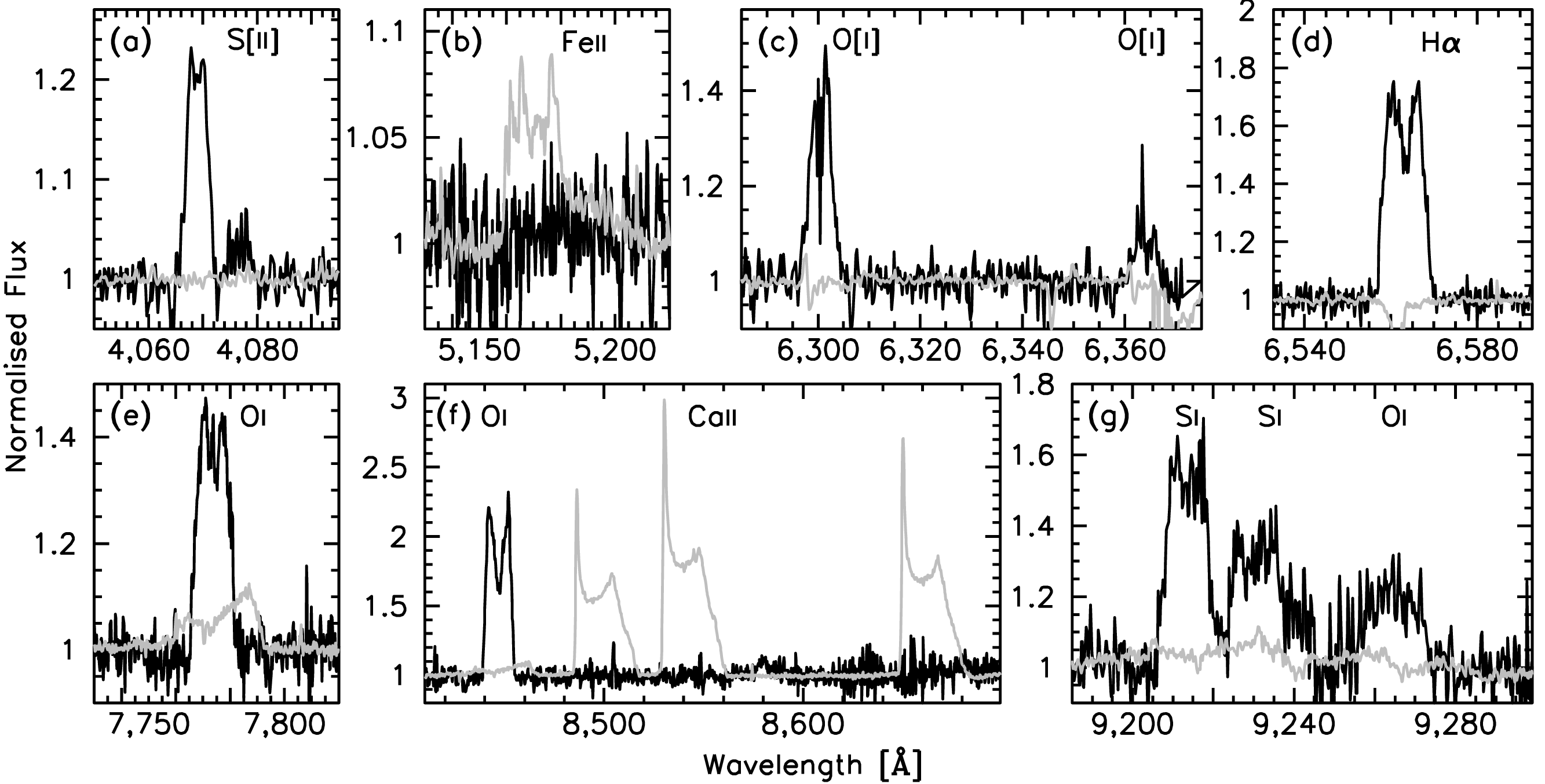}}

\noindent
\textbf{Fig.\,~1.} \textbf{Emission lines from the circumstellar disc at WD\,J0914+1914.}  The X-Shooter spectrum of WD\,J0914+1914 (black) contains strong and broad emission lines of hydrogen, oxygen and sulphur. H$\alpha$ (d) and \Line{O}{i}{8,446\,\AA} (f) are double-peaked, indicating an origin in a circumstellar disc\cite{horne+marsh86-1}. \Line{O}{i}{7,774\,\AA} (e) and the oxygen and sulphur lines near 9,240\,\AA\ (g) are multiplets, resulting in more complex line profiles. The forbidden sulphur and oxygen lines (a, c) have a smaller peak separation, indicating that they are emitted by material extending to larger distances from the white dwarf compared to the other lines. The spectra of the gaseous planetary debris discs detected at several other white dwarfs\cite{gaensickeetal06-3, melisetal12-1} are all dominated by the \Line{Ca}{ii}{8,600\,\AA} triplet (f), with weak additional emission lines of oxygen (e,f), and iron (b), as illustrated by the spectrum of the prototypical system SDSS\,J1228+1040\cite{manseretal16-1} (gray). The striking difference between the two spectra illustrates the different composition of the planetary material~--~gaseous in WD\,J0914+1914, and rocky in SDSS\,J1228+1040. 

\clearpage
\centerline{\includegraphics[width=89mm]{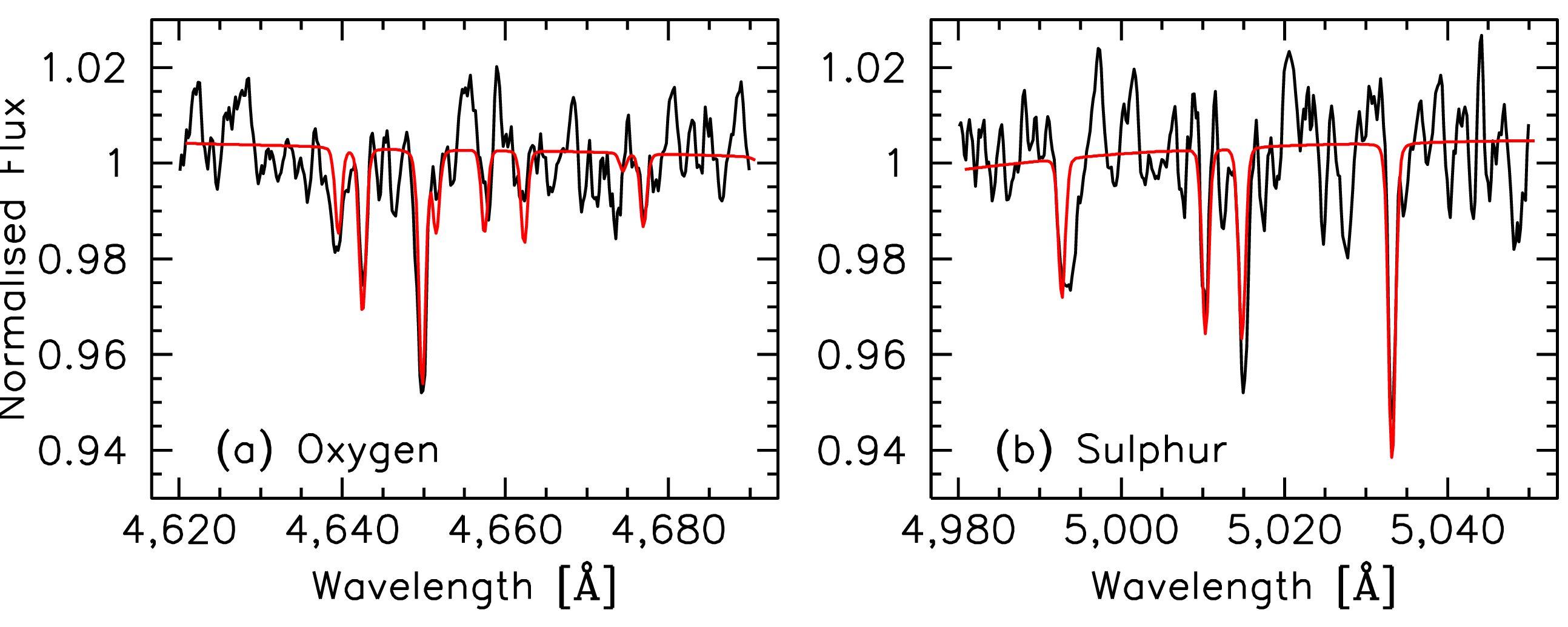}}

\noindent
\textbf{Fig.\,~2.} \textbf{Photospheric oxygen and sulphur lines.} The optical spectrum of WD\,J0914+1914 contains strong photospheric lines of oxygen (a) and sulphur (b), indicating the ongoing accretion from the circumstellar gas disc. A spectral analysis of these lines results in $\log(\mathrm{S/O})=-0.9$ (by number).

\clearpage
\centerline{\includegraphics[width=89mm]{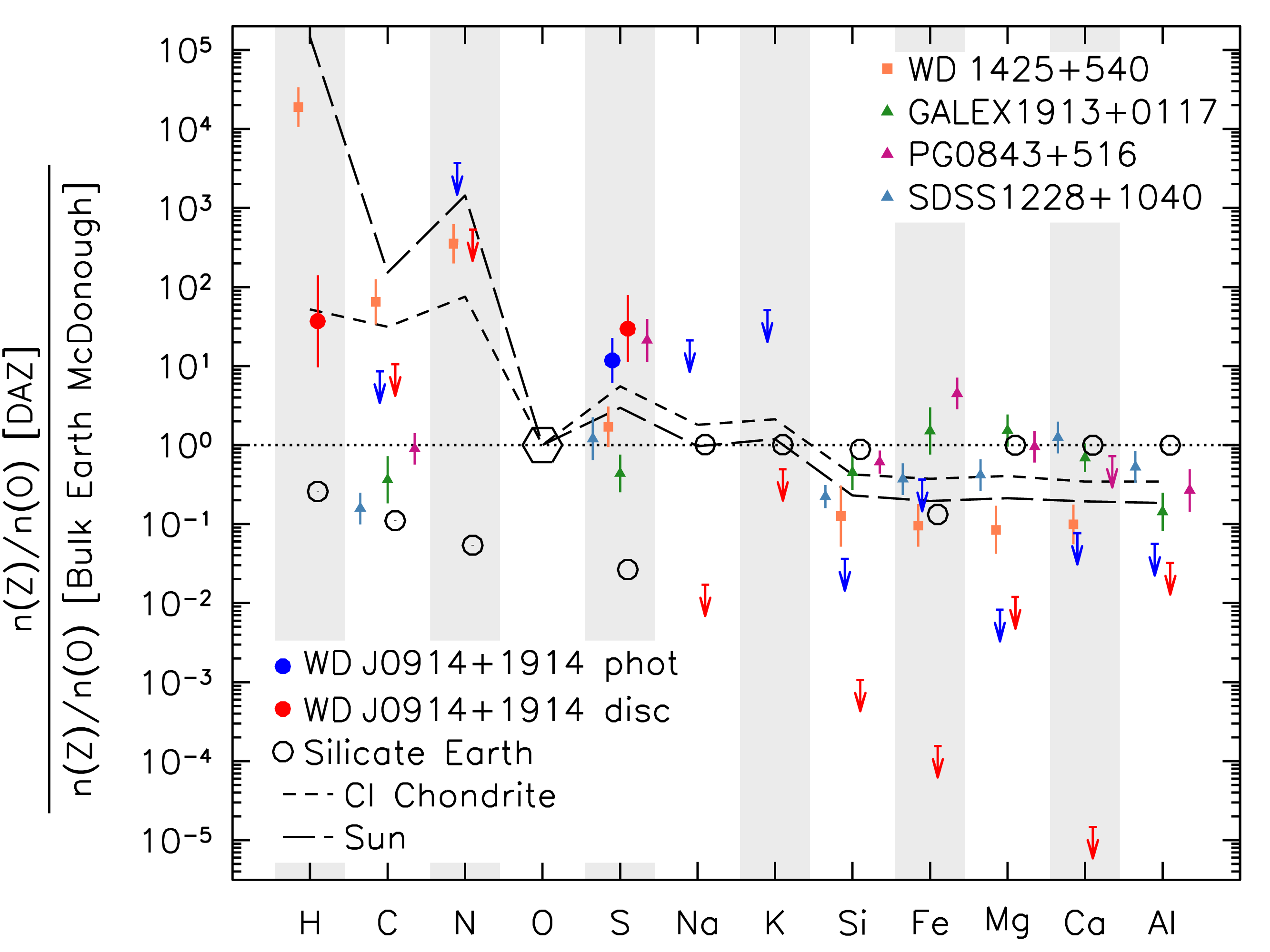}}

\noindent
\textbf{Fig\,3.} \textbf{Abundances of the planetary material at WD\,J0914+1914.} Shown are the number abundances relative to oxygen, normalised to the corresponding ratio for the bulk Earth\cite{mcdonough00-1}, and sorted by condensation temperature. The error bars represent one-sigma uncertainties. The only detected elements are hydrogen (in the circumstellar gas), oxygen and sulphur. Blue dots represent the abundances measured from the analysis of the white dwarf photosphere, red dots represent those derived from the \textsc{Cloudy} photo-ionisation model for the circumstellar gas, and the respective upper limits are shown by downward arrows. Included for comparison are the abundances of the Sun (long dashed lines), CI Chondrites (short dashed lines), three white dwarfs accreting rocky debris \cite{gaensickeetal12-1} (triangles, which scatter closely around the bulk Earth abundances) and the one white dwarf accreting a Kuiper belt-like object\cite{xuetal17-1} (squares, broadly resembling solar abundances).  The material at WD\,J0914+1914 is depleted by orders of magnitude in rock and dust forming elements (Si, Fe, Mg, Ca) with respect to all known minor planetary bodies and stars.

\clearpage

\section*{\large{Methods}}

\subsection{Discovery and follow-up observations.}
%
Two SDSS spectra of WD\,J0914+1914 were taken in November 2005\cite{abazajianetal09-1}, and March 2012\cite{abolfathietal14-1} (see Extended Data Fig.\,1 and Extended Data Table\,1), revealing the H$\alpha$, oxygen and sulphur emission lines. No significant change in the strength of the emission lines is detected between the two epochs. 

We observed WD\,J0914+1914 on 2019 January 12 and 13 using X-Shooter\cite{vernetetal11-1} mounted on UT2 of the Very Large Telescope. X-Shooter is a three-arm spectrograph covering  the extreme blue (UVB, 330--560\,nm), visual (VIS, 560\,nm--1$\mu\mathrm{m}$) and near-infrared (NIR, $1-2.4\,\mu\mathrm{m}$) simultaneously. We obtained ten spectra with 20\,min exposure times each. Given the faintness of the star, $z=19.9$, little signal was expected in the NIR arm, and we therefore used the ``stare'' mode, i.e. avoiding nodding. The data were reduced with the \textsc{REFLEX} package adopting the standard settings and optimising the slit integration limits\cite{freudlingetal13-1}. Finally a weighted average spectrum was computed from the individual UVB and VIS observations. The signal-to-noise ratio of this average spectrum is $\simeq45$ and $\simeq55$ at 4,300\,\AA\ and 7,000\,\AA, respectively.

The X-Shooter spectrum contains the same emission lines detected in the SDSS spectra, plus several additional oxygen and sulphur lines (Fig.\,1). The emission lines are broad and double-peaked, indicating that they originate in a circumstellar disc$^{9,}$\cite{smak81-1}. Also present in the spectrum are multiple strong photospheric absorption lines of oxygen and sulphur, implying ongoing accretion from the disc (Fig.\,2). The detection of the emission lines in the 2019 X-Shooter spectra, and comparison with the 2005 SDSS spectrum places a lower limit of 14 years on the life-time of the disc.

\subsection{Stellar parameters of the white dwarf and its progenitor.}
We measured the atmospheric parameters of WD\,J0914+1914 by fitting pure-hydrogen model spectra\cite{koester10-1} to the two SDSS spectra, which are well flux-calibrated. We used the well-established technique of fitting the Stark-broadened Balmer line profiles\cite{bergeronetal92-1, homeieretal98-1}, which are sensitive to both temperature and gravity.  The total extinction along the line-of-sight towards WD\,J0914+1914 is low, $E(B-V)=0.0305\pm0.0006$\cite{schlafly+finkbeiner11-1}, 
and normalising the Balmer lines prior to the fit effectively removes the effect of extinction. The parameters from the fits to the two SDSS spectra are consistent with each other within the uncertainties, and we take the variance weighted average as the best-fit values (Extended Data Table\,2). Using the cooling  models of \cite{bergeronetal16-1, holberg+bergeron06-1,kowalski+saumon06-1,tremblayetal11-2}, we computed from the effective temperature, $T_\mathrm{eff}=27,743\pm310$\,K and the surface gravity, $\log g=7.85\pm0.06$, a white dwarf mass of $M_\mathrm{wd}=0.56\pm0.03\,\mathrm{M_\odot}$ and a cooling age of $13.3\pm0.5$\,Myr. The quoted uncertainties are only of statistical nature. The magnitude of additional systematic uncertainties can, in principle, be assessed from comparing the results from the spectroscopic fit to a joint analysis of the photometry and parallax of the star\cite{tremblayetal19-1, genest-beaulieu+bergeron19-1}. However, the large parallax uncertainty of WD\,J0914+1914 ($\simeq22$\,per cent) severely limits the precision of the atmospheric parameters derived from such an analysis$^{27}$. The spectrophotometric distance implied by our fit is $\simeq625$\,pc, consistent with the upper limit on the distance based on the \textit{Gaia} parallax\cite{bailer-jonesetal18-1}. As an alternative independent test of our spectroscopic fit, we applied an extinction of $E(B-V)=0.0305$ to the model spectrum with the atmospheric parameters given above, and then scaled that reddened model to the SDSS $r$-band magnitude. We computed a \textit{GALEX} NUV magnitude of 18.07 from this model, which agrees well with the observed value of $18.06\pm0.03$\cite{bianchietal11-1}.

There is still quite some uncertainty in the low-mass end of the initial-to-final mass relation. Using two different relations results in progenitor masses of $\simeq1.0\,\mathrm{M_\odot}$\cite{cummingsetal18-1} and $\simeq1.6\,\mathrm{M_\odot}$\cite{kaliraietal08-1}. The larger value is in closer agreement with many of the earlier works on the initial-to-final mass relation\cite{weidemann00-1, catalanetal08-1, casewelletal09-1, williamsetal09-1}. The main sequence life-times of stars in this mass range are $\simeq2-10$\,Gyr, i.e. the white dwarf cooling age is negligible compared to the total system age.

\subsection{Photospheric abundances.}
Fixing the atmospheric parameters as derived above, $T_\mathrm{eff}=27,743$\,K and $\log g=7.85$, we computed synthetic spectra\cite{koester10-1} for a wide range of abundances of C, N, O, Ne, Na, Mg, Al, Si, P, S, Cl, Ar, K, Ca, Sc, Ti, V, Cr, Mn, Fe, and fitted those models to the average X-Shooter spectrum. The only elements detected in the photosphere are oxygen and sulphur at $\log{\mathrm{(O/H)}}=-3.25\pm0.20$ and $\log{\mathrm{(S/H)}=-4.15\pm0.20}$ (by number), implying $\log{\mathrm{(S/O)}}=-0.9$, which is significantly above the solar value of $-1.57$, though still within the range of stars within the solar neighborhood\cite{hinkeletal14-1}. For all other elements, we derived upper limits (see Extended Data Table\,2). 

Radiative levitation is negligible for oxygen and sulphur at the effective temperature of WD\,J0914+1914 (see Fig.\,2 of \cite{chayeretal95-2}), and therefore the large photospheric abundances of these elements imply ongoing accretion.  Accounting for the diffusion velocities, the photospheric oxygen and sulphur abundances require accretion rates of $\dot{M_\mathrm{S}}=5.5\times10^{8}\,\mathrm{g\,s^{-1}}$ and $\dot{M_\mathrm{O}}=2.7\times10^{9}\,\mathrm{g\,s^{-1}}$, respectively. Several studies argue that the gradient of the mean molecular weight resulting from the accretion of metals into the radiative hydrogen atmospheres of warm white dwarfs drives thermohaline mixing$^{12,}$\cite{dealetal13-1, bauer+bildsten18-1}, which would cause the above rates being underestimated. The most recent studies$^{12,}$\cite{bauer+bildsten18-1} only extend to $T_\mathrm{eff}\simeq20,000$\,K, and we conclude that the combined accretion rate of oxygen and sulphur based on purely diffusive sedimentation provides a lower limit of $\dot M\simeq3.3\times10^9\,\mathrm{g\,s^{-1}}$. The actual rate may be higher by an order of magnitude.

The H$\alpha$ emission line from the circumstellar disc suggests that hydrogen is also accreted onto the white dwarf. However, given that hydrogen is the dominant element in the atmosphere, we are not able to derive the hydrogen fraction in the accreted material. Consequently, the analysis of the photospheric spectrum does not provide a constraint on the contribution of hydrogen to the total accretion rate from the circumstellar disc.

\subsection{Dynamical information on the location of the emitting gas.}
The double-peaked structure of the emission lines arises from the Keplerian motion ($v_\mathrm{K}$) of gas in a disc around the white dwarf, with
\begin{equation}
    v_\mathrm{K}=\sqrt{\frac{GM_\mathrm{wd}}{r}}
\end{equation}
where $G$ is the gravitational constant, and $r$ the distance from the centre of the white dwarf. Hence, the morphology of the emission line profiles provides dynamical information on the location of the emitting gas, with the separation of the double-peaks corresponding to emission from the outer edge of the disc, and the maximum velocity detected in  the line wings corresponding to emission from the inner edge$^{9,}$\cite{smak81-1}. Inspection of the normalised line profiles shows that the morphologies of the individual lines are distinctly different  (see Extended Data Figure\,2). In particular, the double-peak separation of the forbidden [\Ion{S}{ii}] lines is  narrower than that of H$\alpha$ and \Line{O}{i}{8,446\,\AA}, which implies that the region emitting [\Ion{S}{ii}] extends to larger distances from the white dwarf. To estimate the velocity ranges over which the circumstellar gas contributes to the observed emission lines we measured the separation of the double-peaks and the maximum extent of the line wings (full width at zero intensity) of H$\alpha$, \Line{O}{i}{8,846\,\AA} and \Line{[S}{ii]}{4,068\,\AA} (\Line{O}{i}{7,774\,\AA} is a relatively widely spaced triplet, which results in more complex sub-structure of the line profile, and the \Line{[O}{i]}{6,300, 6,343\,\AA} lines are affected by residuals from the oxygen night-sky airglow of the Earth's atmosphere). Whereas the separation of the double-peaks shows a wide range of velocities ($\simeq150\,\mathrm{km\,s^{-1}}$ for \Line{[S}{ii]}{4,068\,\AA}, $\simeq260\,\mathrm{km\,s^{-1}}$ for H$\alpha$ and $\simeq350\,\mathrm{km\,s^{-1}}$ for \Line{O}{i}{8,446\,\AA}), all lines have similar maximum velocities, $\simeq630-650\,\mathrm{km\,s^{-1}}$, implying that they share a common inner radius in the disc.

Because the inclination of our line of sight against the accretion disc is unknown, the semi-major axes of the Keplerian orbits associated with the velocities measured from the emission lines span a wide range. Adopting a white dwarf mass of $M_\mathrm{wd}=0.56\,\mathrm{M_\odot}$, the correspondence between inclination and semi-major axis is illustrated in the Extended Data Fig.\,3. Inclinations $i<5^\circ$ can be excluded as the orbits of the gas would fall inside the white dwarf. For an inclination of $90^\circ$ (edge-on), the inner and outer radii of the gas disc are $\simeq1\,\mathrm{R_\odot}$ and $\simeq10\,\mathrm{R_\odot}$, respectively.

\subsection{A photo-ionisation model for the accretion disc.}
Given the mixture of ionisation species seen in emission (\Ion{H}{i}, \Ion{O}{i}, \Ion{S}{i,ii}), the temperature of the circumstellar gas disc is expected to be in the range $\simeq5,000-10,000$\,K. Whereas mass transfer through the disc will result in some viscous dissipation, the accretion rate  inferred from the photospheric oxygen and sulphur abundances ($\dot M\simeq3.3\times10^9\,\mathrm{g\,s^{-1}}$) cannot provide sufficient heating. This problem has been explored and discussed in detail for the known gaseous debris discs around white dwarfs\cite{hartmannetal11-1}. Instead, photo-ionisation by the intense ultraviolet flux from the white dwarf is extremely efficient at heating the upper layers of the disc\cite{melisetal10-1,kinnear11}. We used the photo-ionisation code \textsc{Cloudy}$^{13}$ to develop a simple model that can provide insight into the geometry and the composition of the circumstellar gas at WD\,J0914+1914. 

\textsc{Cloudy} requires the spectral energy distribution, and luminosity of the ionising source as inputs, for which we computed a white dwarf model spectrum spanning wavelengths from 10\,\AA\ to 3\,$\mu$m with the parameters in the Extended Data Table\,1.  We adopted the solar abundances for the base composition of the circumstellar gas as provided in solar\_GASS10.abn\cite{grevesseetal10-1} within the \textsc{Cloudy} distribution.

The geometry of the irradiation of the disc by the white dwarf can broadly be separated into two regimes, depending on the ratio of the disc height to the radius of the white dwarf. The disc height is given by\cite{franketal02-1}
\begin{equation}
H = \sqrt{k\,\frac{T_{\mathrm{gas}}\,r^{3}}{\mu \, G \, M_{\mathrm{wd}}}}
\end{equation}
where $k$ is the Boltzmann constant, $T_\mathrm{gas}$ the temperature of the gas and $\mu$ the mean molecular weight of the gas.  The mean molecular weight depends on the abundances of the gas (primarily on the mass fractions of hydrogen, oxygen, and sulphur) and on the degree of ionisation, but is not expected to vary much beyond $\mu\sim10-30\,m_\mathrm{p}$, with $m_\mathrm{p}~$ the proton mass. The dominant factor in the above expression is therefore the distance $r$ from the white dwarf, which implies that the disc flares up $\propto r^{3/2}$. 

Near the white dwarf, $r\la 1\mathrm{R_\odot}$, the disc height is small compared to the radius of the white dwarf, and the disc is illuminated from above. However, due to the shallow angle, $\alpha$, of the incident radiation, the effective path length through the gas is much larger than the actual disc height, $H/\sin(\alpha)$. For distances larger than $\simeq 1\mathrm{R_\odot}$, the height of the disc approaches, and eventually exceeds the radius of the white dwarf, and the assumption of a gas shell illuminated by a point source becomes appropriate. We approximated the near case by a gas shell with a distance $r$ from the white dwarf, and a thickness $dr=H/\sin(\alpha)$, and the far case by a gas shell with a distance $r$ from the white dwarf, and $dr$ as free parameter. 

We computed an initial set of \textsc{Cloudy} models, exploring the following free parameters: $r$, the distance from the centre of the white dwarf, $dr$, the extent of the gas layer, $\rho$ the density of the gas, and $\mathrm{H/O}$, the number abundance of hydrogen relative to oxygen.   In these initial models, we fixed $\log\mathrm{(S/O)}=-0.9$, as determined from the analysis of the white dwarf photospheric spectrum. No elements apart from hydrogen, oxygen and sulphur were included in the model at this stage.  The primary input parameter for \textsc{Cloudy} is the hydrogen number density, $N_\mathrm{H}$, which we computed for a given model from the gas density $\rho$, and the H/O and S/O abundance ratios.

The ultraviolet radiation from the white dwarf photo-ionises the upper layers of the circumstellar disc, heating it to $\simeq10,000-20,000$\,K. These layers are optically thin in the continuum, and the cooling of the gas takes place via the emission lines detected in the optical spectrum of WD\,J0914+1914. Deeper layers are essentially neutral, and the observed emission line spectrum does not provide a constraint on the total column density of this neutral material. Within reasonable limits, $\rho$ and $dr$ can be traded off against each other, as both parameters determine the total column density of the gas, and hence the total cross-section for intercepting the ultraviolet photons from the white dwarf.

To assess the quality of the \textsc{Cloudy} models, we computed line flux ratios for all observed emission lines, and compared the values from the synthetic spectrum with those measured from the X-Shooter data:  
\begin{equation}
   Q=\sum_{i=1}^{N_\mathrm{lines}}
        \sum_{j=i+1}^{N_\mathrm{lines}} 
        \frac{F_i^\mathrm{s}/F_j^\mathrm{s}}{F_i^\mathrm{o}/F_j^\mathrm{o}}
        + \frac{F_j^\mathrm{s}/F_i^\mathrm{s}}{F_j^\mathrm{o}/F_i^\mathrm{o}}
        \label{eq:quality}
\end{equation}
where $F^\mathrm{o}$ and $F^\mathrm{s}$ refer to the observed and synthetic line fluxes, respectively. The above function equally penalises models in which the line fluxes are either too large, or too low. 

From the first exploratory models we found that for a solar $\mathrm{O/H}$ ratio, the Balmer lines were always \textit{much} stronger than observed, independent of the exact choice of $r$, $dr$, and $\rho$. Depleting $\log(\mathrm{O/H})\simeq0.29$ resulted in model line flux ratios that were within the right order of magnitude. At close separations from the white dwarf, low densities ($\rho\la10^{-11}\,\mathrm{g\,cm^{-3}}$) are insufficient to cool the gas efficiently, and the resulting line flux ratios are incompatible with the observations. For higher densities, cooling becomes more efficient, and the deeper layers are sufficiently cool to produce significant emission in the \Ion{O}{i} lines. However, the synthetic spectra contain a number of strong lines that are not observed (\Line{O}{i}{3,946\,\AA}, \Line{O}{ii}{4,650\,\AA} and \Line{S}{i}{4,590\,\AA}), and fail to reproduce the line strengths of the observed forbidden lines ([\Ion{O}{i}], [\Ion{S}{ii}]). In conclusion, this first sequence of models indicated that hydrogen is strongly depleted in the disc, and that geometries corresponding to very low inclinations ($i\la20^{\circ}$, see Extended Data Fig.\,3) that would result in inner disc radii with $r\ll1\,\mathrm{R_\odot}$ are incompatible with the observations. 

To find the parameter space that best reproduces the observed line flux ratios we proceeded to compute a grid of \textsc{Cloudy} models with a fixed $dr=0.3\,\mathrm{R_\odot}$, sampling $0.1\,\mathrm{R_\odot}\la r \la10 \mathrm{R_\odot}$ (constrained by the widths of the observed lines, see Extended Data Fig.\,3),  $10^{-9.4}\,\mathrm{g\,cm^{-3}}<\rho<10^{-12}\,\mathrm{g\,cm^{-3}}$ and $\log(\mathrm{O/H})=-0.11$ to $0.89$ and $\log\mathrm{(S/O)}=-1.77$ to $0.23$. The quality of the models in the ($r, \rho$) plane (Extended Data Fig.\,4) illustrates that the best match to the observed line flux ratios is achieved for a location of the gas at $1\,\mathrm{R_\odot}\la r\la 4\,\mathrm{R_\odot}$, and a density of $\rho\simeq10^{-11.3}\,\mathrm{g\,cm^{-3}}$. The synthetic spectra in this parameter range produce line flux ratios that are typically consistent with the observed values within a factor $\simeq2$, and do not result in emission lines that are not detected. Combining the constraints from the \textsc{Cloudy} models with those derived from the profile morphology of the observed emission lines (Extended Data Fig.\,2 \& 3) suggests an inclination of the disc $i\ga50^{\circ}$. 

\subsection{Abundances of the circumstellar disc.}
The best \textsc{Cloudy} models are found for $\log\mathrm{(O/H)}\simeq0.29$ and $\log\mathrm{(S/O)}\simeq-0.5$, with uncertainties of $0.3$\,dex. For comparison, we derived $\log\mathrm{(S/O)}=-0.9$ from the photospheric analysis.  Both measurements agree within a factor $\simeq2.5$. This is the first instance where the composition of the accreted material is consistently determined by two independent measurements, i.e. from the absorption lines within the white dwarf atmosphere, and from the emission lines of the circumstellar gas reservoir.

The fact that the X-Shooter spectrum only contains emission lines of hydrogen, oxygen, and sulphur provides upper limits on the abundances of other elements within the circumstellar gas disc that are typically found in white dwarfs accreting planetary debris. Fixing $r=2.5\,\mathrm{R_\odot}$ and $\rho=10^{-11.2}\,\mathrm{g\,cm^{-3}}$, we  proceeded to add additional elements into the disc model, with their initial abundance set to its solar value. The resulting \textsc{Cloudy} spectra predict strong emission lines for C, N, Na, Mg, Al, Si, K, Ca, and Fe. We re-computed \textsc{Cloudy} models, reducing the abundances until the line strengths in the models were consistent with the non-detection in the X-Shooter spectrum. The upper limits on the abundances of these elements within the circumstellar gas disc are reported in Extended Data Table\,2. Figure\,3 illustrates that these upper limits are much more stringent for Na, Si, Fe, and Ca compared to the limits obtained from the white dwarf photosphere analysis.

\subsection{Emission line profiles from a Keplerian disc.}
The \textsc{Cloudy} model only takes into account the integrated line fluxes. In order to explore how well this model can also reproduce the observed emission line profiles we convolved the \textsc{Cloudy} spectrum from the computed grid that resulted in the best quality (Eq.\,3), corresponding to $r_\mathrm{in}=1.89\,\mathrm{R_\odot}$, $dr=0.3\,\mathrm{R_\odot}$, $\rho=10^{-11.3}\,\mathrm{g\,cm^{-3}}$, $\log{\mathrm{(S/O)}=-0.5}$, and $\log{(\mathrm{H/O})=-0.29}$, with the line profiles of a Keplerian disc. As, at this stage, we are interested in the shape of the line profiles, we normalised the line fluxes of the \textsc{Cloudy} model to those measured from the X-Shooter spectrum, effectively removing the small remaining differences ($\simeq$ factor two, see above) in the absolute line fluxes. We used analytical expressions for the Abel transform\cite{smak81-1}, a power-law index of zero for the radial intensity distribution, and allowed the inner and outer radii of the disc to vary in order to match the observed emission line profiles. Adopting an inclination of the gaseous disc against the line of sight of $i=60^\circ$, the line widths and separations of the double-peaks of H$\alpha$ and \Line{O}{i}{8,446}\,\AA\ are well matched (Extended Data Fig.\,2) by inner disc radii of $r_\mathrm{in}\simeq1.0-1.3\,\mathrm{R_\odot}$ and outer radii of $r_\mathrm{out}\simeq3.0-3.3\,\mathrm{R_\odot}$. The more complex structure of the \Line{O}{i}{7,774\,\AA} multiplet is also reasonably well reproduced by the same range of $r_\mathrm{in}$ and $r_\mathrm{out}$. In contrast, \Line{[S}{ii]}{4,068\,\AA} requires $r_\mathrm{in}\simeq1.0-1.3\,\mathrm{R}_\odot$ and $r_\mathrm{out}\simeq8-10\,\mathrm{R}_\odot$, and the two forbidden \Ion{[O}{i]} lines also imply similarly large outer radii, even if their double-peaks are not well resolved due to the residuals of the sky background subtraction. The larger outer disc radii implied by the line profiles of the forbidden lines confirm the simple estimates we made above (see Extended Data Fig.\,3). 

While the synthetic line profiles of an axially symmetric disc reproduce the X-Shooter data relatively well, there is a noticeable difference in the shape of the central depression of \Line{O}{i}{8,446\,\AA}, with the observations showing a deeper V-shape compared to the U-shape of the model line profile. Similar differences have been observed in the Balmer lines from accretion discs in white dwarf binaries, and have been interpreted as optical depth effects\cite{marsh87-1}. We also note that matching the observed width of the H$\alpha$ double-peaks requires a small amount of additional intrinsic broadening, which could be the result of Stark broadening within the disc\cite{marsh87-1}. 

We conclude that despite our model for the circumstellar gas disc being relatively simple (based on a constant  density both in radius and vertical extent of the disc), the overall agreement in both the emerging fluxes and the profile morphology of the emission lines is remarkably good, resulting in a consistent set of parameters both in terms of the geometric location of the gas, and its composition. The reality will have a more complex geometry as well as density gradients. However, including that complexity in the model by introducing additional free parameters is unlikely to provide deeper physical insight.

\subsection{Ruling out a stellar / sub-stellar companion.}
%
The initial classification of WD\,J0914+1914 suggested it to be a cataclysmic variable (CV), i.e. a short-period binary containing a white dwarf accreting from a Roche-lobe filling low-mass companion. Whereas the double-peaked morphology of the emission lines confirms the presence of a circumstellar gas disc, CVs typically have much stronger Balmer (and often helium) lines\cite{szkodyetal07-2, skodyetal10-1, breedtetal14-1, thorstensenetal16-1}, and no example of a CV with a white dwarf as hot as $\simeq28\,000$\,K dominating the optical spectrum is known\cite{gaensickeetal09-2, palaetal17-1}. 

Another class of systems with similar spectroscopic appearance as WD\,J0914+1914 are detached short-period post-common envelope binaries (PCEBs), i.e. white dwarf binaries with low-mass companions, where H$\alpha$ emission from the companion star is commonly detected\cite{hillwigetal00-1, kawkaetal00-1, odonoghueetal03-1}. In PCEBs containing hot white dwarfs, emission lines of calcium and iron originate from the intense irradiation of the companion\cite{schmidtetal95-3, aungwerojwitetal07-1}, which are not observed in WD\,J0914+1914. The emission lines in PCEBs are narrow and single-peaked, and trace the orbital motion of the companion star, with typical periods of hours and radial velocity amplitudes of several $100\,\mathrm{km\,s^{-1}}$\cite{maxtedetal06-1,parsonsetal17-1}. The double-peaked shape of the emission lines in WD\,J0914+1914 already rules out an origin from an irradiated low-mass companion. Moreover, their velocity variation is $\la20\,\mathrm{km\,s^{-1}}$, much lower than observed in any of the known PCEBs\cite{parsonsetal17-1}.

We measured the radial velocity of the white dwarf using ten of the strongest sulphur absorption lines in the X-Shooter UVB spectra. We fixed the relative wavelengths of these lines to their laboratory values, and their width to 1\,\AA, roughly matching the spectral resolving power, leaving only the depths of the lines, and the white dwarf radial velocity as free parameters. We find a mean white dwarf velocity of $-47\,\mathrm{km\,s^{-1}}$ and an average statistical uncertainty of the individual velocity measurements of $\simeq4.0\,\mathrm{km\,s^{-1}}$. In addition, there is a systematic uncertainty arising from imperfections in centring the star in the slit and the instrument model accounting for flexure. We measured this systematic uncertainty from the interstellar Ca\,K line to be $\simeq3.7\,\mathrm{km\,s^{-1}}$, and added it in quadrature to the statistical uncertainties, resulting in a total uncertainty of the individual radial velocity measurements of $\simeq5.5\,\mathrm{km\,s^{-1}}$. The radial velocities of WD\,J0914+1914 are consistent with a constant value, i.e. the reduced $\chi^2$ with respect to the mean is $\chi_\mathrm{red}^2=0.95$. We conclude that we do not detect a radial velocity variation of the white dwarf, with an upper limit on its radial velocity amplitude of $K_\mathrm{wd}\la3\,\mathrm{km\,s^{-1}}$. For the typical periods of PCEBs, $\simeq2\,\mathrm{h}-1\,\mathrm{d}$\cite{nebotetal11-1}, brown dwarf companions are ruled out. In the period range for the mass donating object suggested by our analysis (see below), $\simeq8-10$\,d, companions with $M\ga30\,\mathrm{M_{Jup}}$ are ruled out. 

Furthermore, a stellar companion would result in an infrared excess with respect to an isolated white dwarf. The location of WD\,J0914+1914 has been covered by the UKIRT Hemisphere Survey (UHS\cite{dyeetal18-1}) in the $J$-band. WD\,J0914+1914 is not detected at the $J=19.6$ ($5\sigma$) magnitude limit of UHS. The white dwarf alone has $J=19.65$, computed from the synthetic spectrum. Using absolute $J$-band magnitudes of M-dwarfs and L-type brown dwarfs\cite{hoardetal07-1}, and a conservative upper limit on the distance of $d=631$\,pc\cite{bailer-jonesetal18-1}, the non-detection of WD\,J0914+1914 in UHS excludes the presence of a companion earlier than an L5 brown dwarf.

The forbidden oxygen and sulphur lines detected in the spectrum of WD\,J0914+1914 have not been observed in any accreting or detached white dwarf binary.  Accretion from the wind of a low-mass companion does result in photospheric metal contamination in these binaries\cite{debes06-1}, however, their abundances derived from spectroscopic analysis are broadly consistent with solar abundances of the accreted material, with strong absorption lines of calcium, iron, magnesium and silicon$^{14,16,}$\cite{tappertetal11-1}.

We conclude from the analysis of the observations that WD\,J0914+1914 is a white dwarf accreting from a circumstellar gas disc with extremely non-solar abundances, and that the origin of the circumstellar disc is not a stellar or brown-dwarf companion.

\subsection{Photo-evaporation vs Roche-lobe overflow.}
%
The disc size provides constraints on the location of the planet. We assume that the outer radius of disc is traced by the forbidden [\Ion{S}{ii}] and [\Ion{O}{i}] lines, $r_\mathrm{out}\simeq10\,\mathrm{R}_\odot$, and that this radius corresponds to the maximum size of the accretion disc allowed by tidal forces, which is approximately 90~per cent of the white dwarf's Roche-lobe, i.e. $r_{\mathrm{out}}\sim0.9 R_{\mathrm{L_{wd}}}$\cite{franketal02-1}. For a given mass of the planet, assuming a circular orbit, and using standard formula for the Roche-lobe radius\cite{eggleton83-1}, this expression allows an estimate of semi-major axis of the planet's orbit. For Neptune to Jupiter-mass giant planets this implies semi-major axes of $\simeq14-16\,\mathrm{R}_\odot$ and orbital periods of a $\simeq8-10$ days. We envisage two scenarios in which WD\,J0914+1914 could accrete from a planet on a close orbit, i.e. either via mass loss driven by the intense extreme ultraviolet (EUV) luminosity of the white dwarf, or via Roche-lobe overflow. Both alternatives are discussed in detail below.

\textit{Photo-evaporation.} 
EUV radiation is known to drive atmospheric mass loss from giant planets in close orbits around their host stars. This hydrodynamic escape is the result of the ionisation of hydrogen. In the absence of efficient cooling mechanisms, no hydrostatic solution exists for the atmosphere of a planet subject to intense irradiation, leading to the formation of a trans-sonic flow\cite{owen18-1}. Drag forces in this outflow cause heavier elements to be carried with the escaping hydrogen. The detection of Ly$\alpha$ absorption from atomic hydrogen located outside  the Roche-lobe of the transiting planet HD\,209458b clearly demonstrated the escape of atmospheric material from the planet, and provided the first direct evidence for the evaporation of exo-planets\cite{vidal-madjaretal03-1}. Subsequent Ly$\alpha$ transits were detected in a number of other systems, including the hot Jupiter HD\,189733b \cite{lecavelierdesetangsetal10-1} and the close-in Neptune mass planet GJ436b$^{20,}$\cite{kulowetal14-1, lavieetal17-1}. 

In addition to these Ly$\alpha$ transit observations, heavier elements in the extended atmospheres of transiting planets were detected in ultraviolet and X-ray transit spectroscopy\cite{vidal-madjaretal13-1, ben-jaffel+ballester13-1, poppenhaegeretal13-1}, showing that the atmospheric escape must be driven by a hydrodynamic process.  Apart from the observational detection in a number of individual systems, hydrodynamic escape is thought to play a crucial role in shaping the properties of the population of close-in exo-planets\cite{owen18-1}, resulting in the nearly complete absence of Neptune mass planets with orbital periods of a few days (the warm Neptune desert) as well as the dearth of low-mass planets with $1.5-2$ Earth radii (the evaporation valley). 

To test the plausibility of hydrodynamic escape for the planet around WD\,J0914+1914 we determined the EUV flux of the white dwarf and estimated the corresponding evaporation rates using scaling laws derived from detailed hydrodynamic models\cite{murray-clayetal09-1, owen18-1}. The incident EUV flux at the position of the planet was obtained by integrating white dwarf model spectra\cite{koester10-1} from 10 to 912\,\AA.

Trace metals in the photosphere of WD\,J0914+1914, resulting from the accretion of planetary material, may block some of the EUV emission\cite{chayeretal95-1}. To evaluate the significance of EUV line blanketing, we computed three white dwarf models, fixing the effective temperature and surface gravity to the values determined from photospheric analysis (Extended Data Table\,1): (1) a pure-hydrogen model, (2) a hydrogen model with oxygen and sulphur at the photospheric abundances (Extended Data Table\,2), and (3) a hydrogen model including in addition C, N, Na, Mg, Al, Si,  P, Cl, Ar,  K, Ca, Ti, V, Mn, Fe using the lower of the two upper limits on their abundances (photospheric or disc, Extended Data Table\,2) and solar abundances for those elements without meaningful upper limits. We find very small variations of the EUV flux (less than 10 per cent) between the three models, i.e. the amount of metal pollution is insufficient to cause significant line blanketing. Below, we use the model (2), including photospheric sulphur and oxygen. The EUV flux incident upon the planet is shown as a function of orbital separation in the upper panel of Extended Data Fig.\,5. 

The EUV luminosity of WD\,J0914+1914 is comparable to that of T\,Tauri stars which are assumed to efficiently evaporate the atmospheres of their young giant planets. In particular, the atmospheres of Neptunes at separations below 0.1\,au (roughly the outer border of the warm Neptune desert) are supposed to lose significant parts of their atmospheres during these early stages. Analogously, the large EUV luminosity of WD\,J0914+1914 implies that hydrodynamic escape is unavoidable for any planet with a hydrogen-rich atmosphere and a semi-major axis $\la200\,\mathrm{R}_\odot$. 

For large EUV fluxes, hydrodynamic escape can be in the energy limited or the recombination limited regime. Hydrodynamic mass loss scales proportional to the EUV irradiation in the energy-limited regime, and scales with the square root of the EUV irradiation in the recombination-limited regime.

For a Jupiter-mass planet, the transition between both regimes is usually assumed\cite{murray-clayetal09-1} to occur at $10,000\,\mathrm{erg\,cm^{-2}s^{-1}}$ but can vary depending on the mass and the radius of the planet across a wide range of EUV fluxes\cite{owen+alvarez16-1},  $\simeq1,000-100,000\,\mathrm{erg\,cm^{-2}s^{-1}}$. Given that we  currently do not know the mass and radius of the planet at WD\,J0914+1914, we assume $10,000\,\mathrm{erg\,cm^{-2}s^{-1}}$ for the transition. Consequently, the mass loss rates we calculate below should be considered as an order-of-magnitude estimate. For the mass loss rate in the energy limited regime of irradiated giant planets we used\cite{murray-clayetal09-1}: 
\begin{equation}
    \dot{M}=\frac{\epsilon \pi F_{\mathrm{EUV}} R^{3}_{\mathrm{P}}}{GM_{\mathrm{P}}K(\xi)} 
\end{equation}
where $R_{\mathrm{P}}$ and $M_{\mathrm{P}}$ are the radius and the mass of the planet, $F_{\mathrm{EUV}}$ is the incident EUV flux, and $\epsilon$ the efficiency of using the incident energy, which we set to $\epsilon=0.3$ \cite{murray-clayetal09-1,owen+alvarez16-1}. At close orbital separations, where the Roche-lobe ($R_{\mathrm{L_P}}$) and planet radius become comparable, mass loss is enhanced. This is accounted for by the correction term $K(\xi=R_\mathrm{L_P}/R_\mathrm{P})$ for which we used equation 17 from\cite{erkaevetal07-1}. The mass loss rate driven by the strong EUV irradiation from WD\,J0914+1914 is shown in the bottom panel of the Extended Data Fig.\,5. The $K$-term is responsible for the steep increase of $\dot M$ towards the smallest separations. For the estimated location of the planet, ($\simeq14-16\,\mathrm{R_\odot}$, grey shaded region) we obtain a mass loss rate of $\simeq5\times 10^{11}\,\mathrm{g\,s^{-1}}$, depending only mildly on the planet mass. At that distance from the planet, the outflow velocities that are required to reach the Roche-lobe of the planet are far smaller than the velocity required to escape the gravitational potential of the white dwarf, and consequently the evaporated material will fall towards the white dwarf. 

Hydrogen is likely the dominant species in the planet's atmosphere, driving the hydrodynamic escape, and is hence also expected to be the most abundant element in the circumstellar disc at WD\,J0914+1914. However, the weakness of H$\alpha$ in the X-Shooter spectrum, compared to the emission lines of oxygen and sulphur already suggests a significant depletion of hydrogen in the disc with respect to solar abundances, which we quantitatively confirmed with the \textsc{Cloudy} photoionisation models ($\log({\mathrm{O/H}})\simeq0.29$, compared to the solar ratio of $-3.31$). 

Motivated by these results, we explored the effect of forces other than gravity upon the material escaping the  atmosphere of the planet. Being a hot white dwarf, WD\,J0914+1914 is not only bright in the EUV but also in the far-ultraviolet, where radiation pressure from Ly$\alpha$ photons can transfer momentum to the evaporated hydrogen. This effect is well-studied in the solar system, where the radiation pressure acting upon neutral hydrogen atoms within the heliosphere is proportional to the total flux in the solar Ly$\alpha$ emission line. The relative importance of radiation pressure is usually expressed as $\mu$, the ratio between the force related to radiation pressure and gravity, which is very accurately known for the Sun. The effect of Ly$\alpha$ radiation pressure on the motion of interplanetary neutral hydrogen has been measured\cite{schwadronetal13-1} and depending on the solar cycle $\mu$ varies between $\simeq0.8$ during the minimum of solar activity and $\simeq1.8$ during the maximum \cite{bzowskietal13-1}. Heavier species than hydrogen are much less affected by radiation pressure. 

To establish the importance of radiation pressure on the material accreting onto WD\,J0914+1914 we compared the Ly$\alpha$ flux of the white dwarf with that of the Sun. For that purpose, we retrieved the high-resolution far-ultraviolet spectra of the Sun obtained with the SORCE SOLSTICE instrument\cite{mcclintocketal05-1}. Despite the fact that the white dwarf spectrum shows Ly$\alpha$ in absorption, whereas it is in emission in the spectrum of the Sun, the flux in the very core of Ly$\alpha$ of WD\,J0914+1914 is comparable to that of the Sun (Extended Data Fig.\,7), a simple consequence of the high effective temperature of the white dwarf. Moreover, the flux in WD\,J0914+1914 rapidly increases outside the core of Ly$\alpha$, whereas it drops in the Sun, and as $M_\mathrm{wd}<\mathrm{M_\odot}$, $\mu\gg1$. We conclude that this provides a natural explanation for the low abundance of hydrogen in the circumstellar disc at WD\,J0914+1914: the strong radiation pressure from the Ly$\alpha$ photons of WD\,J0914+1914 efficiently inhibits the inflow of hydrogen. This radiation-pressure driven hydrogen depletion of the material flowing towards  the white dwarf results in an accretion rate onto WD\,J0914+1914 that is significantly smaller than the estimated mass loss rate ($\simeq5\times10^{11}\,\mathrm{g\,s^{-1}}$, see Extended Data Fig.\,5). Given that the mass loss rate is an order-of-magnitude estimate only, we conclude that hydrodynamic escape and subsequent accretion of the heavier elements that are dragged by the escaping hydrogen, provides a consistent explanation of our observations.

\textit{Roche-lobe overflow.} 
An alternative possibility for accretion from a  giant planet onto  WD\,J0914+1914 is Roche-lobe overflow which can be significantly increased by tidal heating \cite{valsecchietal15-1}. However, the scenario of Roche-lobe overflow appears to be extremely unlikely for WD\,J0914+1914 for several reasons. Crucially, the observed emission lines are best reproduced by a circumstellar disc extending up to $\simeq10\,\mathrm{R_\odot}$, which implies that the planet must be located at $>10\,\mathrm{R}_\odot$ from WD\,J0914+1914. Even a Jupiter mass planet would have to be substantially inflated (to $\simeq8$ Jupiter radii) to fill its Roche-lobe at such a large orbital separation. More generally, using an empirical mass-radius relation for  giant planets\cite{bashietal17-1}, we find Roche-lobe overflow should occur at separations of $1-2\,\mathrm{R_\odot}$, clearly incompatible with the derived disc size. Furthermore, the mass transfer rates expected from a Roche-lobe overflow configuration exceed the value we derived from the photospheric analysis by several orders of magnitude. We conclude that Roche-lobe overflow is incompatible with the observational characteristics of WD\,J0914+1914.

\subsection{Common envelope evolution vs planet-planet scattering.}
Whereas the observational evidence for a  giant planet in a close-in orbit around WD\,J0914+1914 is compelling, it is clear that a planet with an initial semi-major axis of a few ten $\mathrm{R}_\odot$ would not have survived the red giant branch evolution of the white dwarf progenitor. The physical mechanism that migrated the planet from several au onto its current orbit is open to some speculation. 

One possibility is common envelope evolution. At the onset of a common envelope, dynamically unstable mass transfer starts from the giant star onto the secondary object, in our case the planet. The time-scale for this unstable mass transfer becomes quickly shorter than the thermal time-scale of the planet and, as a result, a common envelope forms around the planet and the core of the giant star, the future white dwarf. This common envelope is expelled at the expense of orbital energy, i.e. the planet spirals inward. 

Common envelope evolution is known to produce binaries containing white dwarfs with stellar\cite{nebotetal11-1} and sub-stellar\cite{maxtedetal06-1, farihietal17-1} companions and orbital periods in the range of hours to days. In fact, common envelope evolution involving planetary mass objects has been suggested as a possible scenario for the formation of low-mass white dwarfs without a detectable stellar companion$^{25}$. As the mass of the planet, $M_{\mathrm{P}}$, is much smaller than the white dwarf mass, the final separation after common envelope evolution can be written as$^{25}$: 
\begin{equation}
    a_f=\frac{\alpha_{\mathrm{CE}} \lambda}{2}\frac{M_{\mathrm{core}} M_{\mathrm{P}}}{M\,M_{\mathrm{env}}} R_{\mathrm{G}}
\end{equation}
where $R_{\mathrm{G}}$ is the radius of the giant star at the beginning of the inspiral phase, $\alpha_{\mathrm{CE}}$ the common envelope efficiency, $\lambda$ the binding energy parameter, $M$ the mass of the giant star which can be separated into the core mass (mass of the future white dwarf, $M_{\mathrm{core}}$) and the envelope mass ($M_{\mathrm{env}}$). The latter is going to be expelled during the process. During common envelope evolution, the planet will move inside the envelope of the giant star and is likely to be completely evaporated. Whether this happens, and at what separation, depends on the temperature structure of the giant star envelope which can be approximated by$^{25}$
\begin{equation}
    T \simeq 1.78 \times 10^6 \times (r/\mathrm{R_\odot})^{-0.85} \mathrm{K}. 
\end{equation}
The radius at which evaporation of the planet occurs can then be estimated by equating the local sound speed in the envelope and the escape velocity of the planet \cite{soker98-1}.

The above approach has been previously used to estimate the outcome of common envelope evolution involving planetary mass companions using constant values for $\alpha_{\mathrm{CE}}$ and $\lambda$$^{25}$. Throughout the last two decades, however, new constraints on the common envelope efficiency $\alpha_{\mathrm{CE}}$ have been obtained and algorithms have been developed that calculate the binding energy parameter $\lambda$, which has been found to sensitively depend on the mass and radius of the giant star, in particular if recombination energy stored in the envelope is assumed to contribute to expelling the envelope  \cite{dewi+tauris00-1,zorotovicetal11-2}. The contributions from recombination energy are usually parametrised with a second efficiency parameter $\alpha_{\mathrm{rec}}$.  

We calculated the possible outcome of common envelope evolution involving a planetary mass companion taking into account these recent developments. We used the \textsc{BSE} code\cite{hurleyetal02-1} to compute the evolution of main sequence stars in the range of $1-8\,\mathrm{M_\odot}$ and determined the binding energy parameter for all core masses close to the mass of WD\,J0914+1914, i.e. we accepted all masses in the range $0.55-0.57\,\mathrm{M_\odot}$. We then used Eq.\,5 to determine the final separation for planet masses ranging from super-earths to the brown dwarf limit. For the planet to survive, the final separation must be sufficiently large that the planet does not evaporate in the red giant envelope, and does not fill its Roche-radius. The latter was calculated from the planet and white dwarf mass and assuming an empirical mass-radius relation for  giant planets\cite{bashietal17-1}.

The results derived from these calculations are illustrated in Extended Data Fig.\,7 for two different values of the common envelope parameters. First, we used the strict upper limit for the contributions from orbital energy and recombination energy, i.e. we assumed that both energies fully contribute to expelling the envelope ($\alpha_{\mathrm{CE}}=\alpha_{\mathrm{rec}}=1.0$). These calculations provide a stringent upper limit for the final separation (shown as the dashed line in Extended Data Fig.\,7). More realistic are smaller values for both efficiencies, e.g. observations of white dwarf binaries with M-dwarf stellar companions can be reproduced if $\alpha_{\mathrm{CE}}=\alpha_{\mathrm{rec}}=0.25$\cite{zorotovicetal10-1}, for which the predicted final separations fall below the solid black line in Extended Data Fig.\,7.

The most important conclusion drawn from inspection of Extended Data Fig.\,7 is that planets with masses smaller than $\simeq1\,M_{\mathrm{Jup}}$ cannot survive common envelope evolution, whereas planets in the mass range of $\sim1-13 M_{\mathrm{Jup}}$ could end up with orbital separations consistent with the estimated location of the planet around WD\,J0914+1914 ($14-16\,\mathrm{R_{\odot}}$). In the latter case the initial planet-star separation must have been $\simeq1.5-5$\,au (depending on the planet mass) at the onset of mass transfer from the giant star onto the planet, when the giant star was close to the end of its AGB evolution, the binding energy of envelope was smallest and the contributions from recombination energy were largest. Planet population synthesis models predict the fraction of giant planets to increase with stellar mass in agreement with recent observational studies\cite{borgnietetal19-1}. Most of the white dwarfs in the Galaxy descend from A/F-type stars, and hence their progenitors are likely to have had rich planetary systems. Given that WD\,J0914+1914 is unique among $\simeq7,000$ white dwarfs with similar cooling ages observed by the SDSS, common envelope evolution can plausibly explain the close-in orbit of the planet at WD\,J0914+1914, but requires it to be more massive than Jupiter.

An alternative scenario explaining the existence of a  giant planet in a close-in orbit around WD\,J0914+1914 is planet-planet scattering. Dynamical studies have shown that closely packed planetary systems which remain stable and ordered on the main sequence can become unpacked when the star evolves into a white dwarf\cite{veras+gaensicke15-1}. As a consequence of this unpacking, inward incursions of planets can occur throughout the entire white dwarf cooling track for basically all types of planetary masses, ranging from Earth-like objects to giant planets.  These inward incursions of planets on largely eccentric orbits will generate strong tidal forces that can circularise the planetary orbit. Planet-planet scattering therefore represents an alternative explanation for the close planet being evaporated by WD\,J0914+1914, and  also works for planet masses lower than the limit for common envelope evolution ($\simeq1\,\mathrm{M_{Jup}}$).

The large abundance of sulphur in the circumstellar disc at WD\,J0914+1914 might indicate a planetary mass closer to Neptune/Uranus as the fraction of heavier elements is thought to increase with decreasing planetary mass\cite{thorngren+fortney19-1}, which would point towards planet-planet scattering causing the inward migration of the planet at WD\,J0914+1914. However, given the high mass loss rates expected from hydrodynamic escape, we cannot exclude a more massive planet. We therefore conclude that given the currently available observational constraints, both planet-planet scattering as well as common envelope evolution are plausible explanations for the existence of the planet in close orbit around WD\,J0914+1914.

Additional constraints on the composition of the accreted material from ultraviolet spectroscopy of WD\,J0914+1914, as well as including tidal effects into N-body simulations of the evolution of  planetary systems around white dwarfs, have the potential to distinguish between the two scenarios.  

\subsection{The past and future of the planet around WD\,J0914+1914.}
As the evolution of white dwarfs is relatively well understood and primarily consists of thermal heat loss through the non-degenerate envelope, and the consequent contraction of this envelope\cite{fontaineetal01-1}, we can predict the incident flux, and with it the evaporation rate of the planet at WD\,J0914+1914, and the resulting accretion rate onto the white dwarf, as a function of time. To that end we computed a small grid of white dwarf model spectra covering effective temperatures ranging from 80,000\,K to 10,000\,K, for the surface gravities corresponding to $M_\mathrm{wd}=0.56\,\mathrm{M}_\odot$ at each temperature.  Integrating the EUV fluxes of these model spectra, we then used Eq.\,4 to estimate the  mass loss rate as a function of effective temperature and cooling age (see Extended Data Fig.\,8). As expected, the mass loss rate decreases with time, particularly once the incident flux on the planet drops below $10,000\,\mathrm{erg\,cm^{-2}s^{-1}}$, when mass loss becomes directly proportional to the EUV flux. We estimate that accretion of the evaporating material will become undetectable via photospheric metal contamination$^3$ once the white dwarf has cooled to $\simeq12,000$\,K, corresponding to a cooling age of $\simeq350$\,Myrs, when the mass loss rate drops below $10^6\,\mathrm{g\,s^{-1}}$.

We estimate the total mass loss due to evaporation of the planetary atmosphere by integrating the mass loss rate over the cooling age of the white dwarf, and assuming that the planet reached its current orbit soon after the formation of the white dwarf. The resulting total mass loss is $\simeq0.002$ Jupiter masses, or $\simeq0.04$ Neptune masses. Thus, hydrodynamic escape will not significantly change the structure of the  giant planet around WD\,J0914+1914. 

\clearpage

\newcounter{mybibstartvalue}
\setcounter{mybibstartvalue}{29}

\xpatchcmd{\thebibliography}{
\usecounter{enumiv}}{\usecounter{enumiv}
\setcounter{enumiv}{\value{mybibstartvalue}}}{}{}

\clearpage
\centerline{\includegraphics[width=120mm]{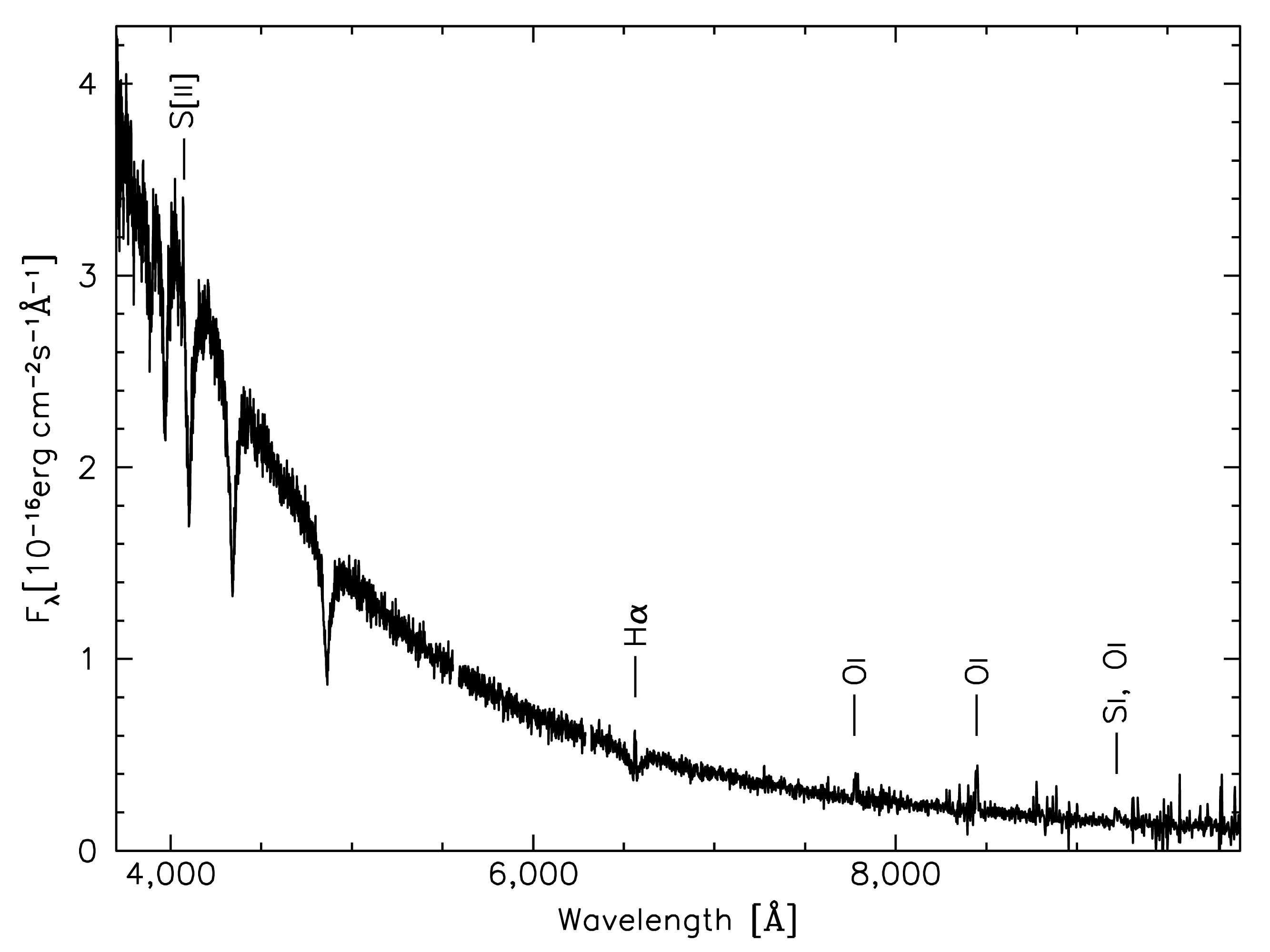}}

\noindent
\textbf{Extended Data Figure\,1.} \textbf{Identification spectrum of WD\,J0914+1914.} The unusual nature of WD\,J0914+1914 was identified from its optical spectrum within the SDSS Data Release~14. The H$\alpha$, \Line{O}{i}{7,774\,\AA} and \Line{O}{i}{8,446\,\AA} lines are clearly detected, \Line{S}{[ii]}{4,068\,\AA} and a blend of \Ion{S}{i} and \Ion{O}{i} near 9,240\,\AA\ are present near the noise level. 

\clearpage
\centerline{\includegraphics[width=120mm]{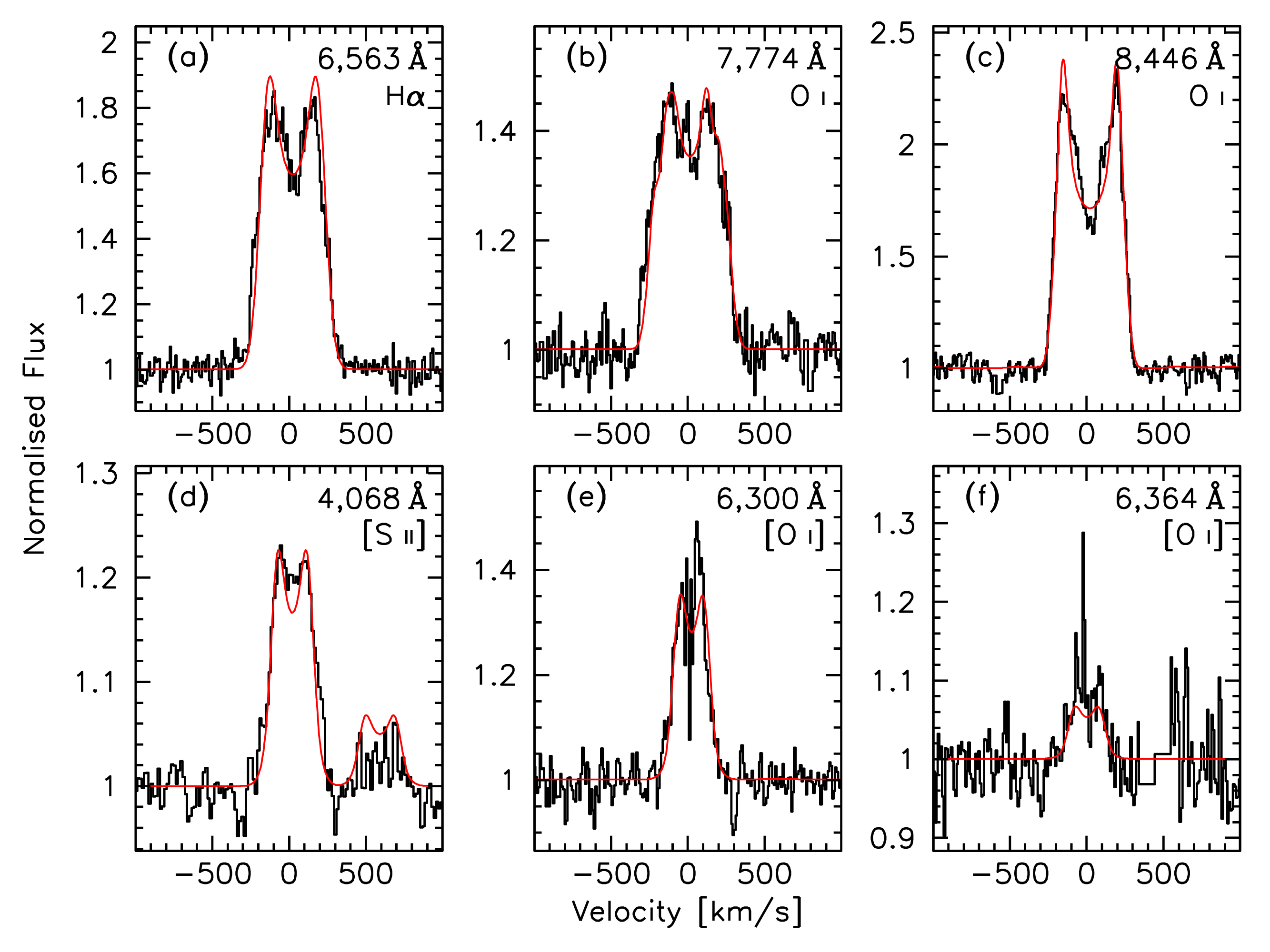}}

\noindent
\textbf{Extended Data Figure\,2.} \textbf{Emission lines from a Keplerian disc.} The double-peaked emission lines of hydrogen (a), oxygen (b, c, e, f) and sulphur (d) detected in the optical spectrum of WD\,J0914+1914 originate in a gaseous circumstellar disc.  Shown in red are synthetic disc profiles computed by convolving the \textsc{Cloudy} model that best matches the observed line flux ratios with the broadening function of a Keplerian disc. Adopting an inclination of $i=60^\circ$, the widths and double-peak separations of the H$\alpha$ (a) and \Line{O}{i}{8,446\,\AA} (c) lines are well reproduced for inner and outer disc radii of $r_\mathrm{in}\simeq1.0-1.3\,\mathrm{R}_\odot$ and $r_\mathrm{out}\simeq2.8-3.3\,\mathrm{R}_\odot$, respectively, consistent with the results from the \textsc{Cloudy} models (see Extended Data Fig.\,4). The emission of \Line{[S}{ii]}{4,068\,\AA} (d) extends from $\simeq1\,\mathrm{R_\odot}-10\,\mathrm{R_\odot}$. The V-shaped central depression of the \Line{O}{i}{8,446\,\AA} (c) line suggests that the line is optically thick.

\clearpage
\centerline{\includegraphics[width=120mm]{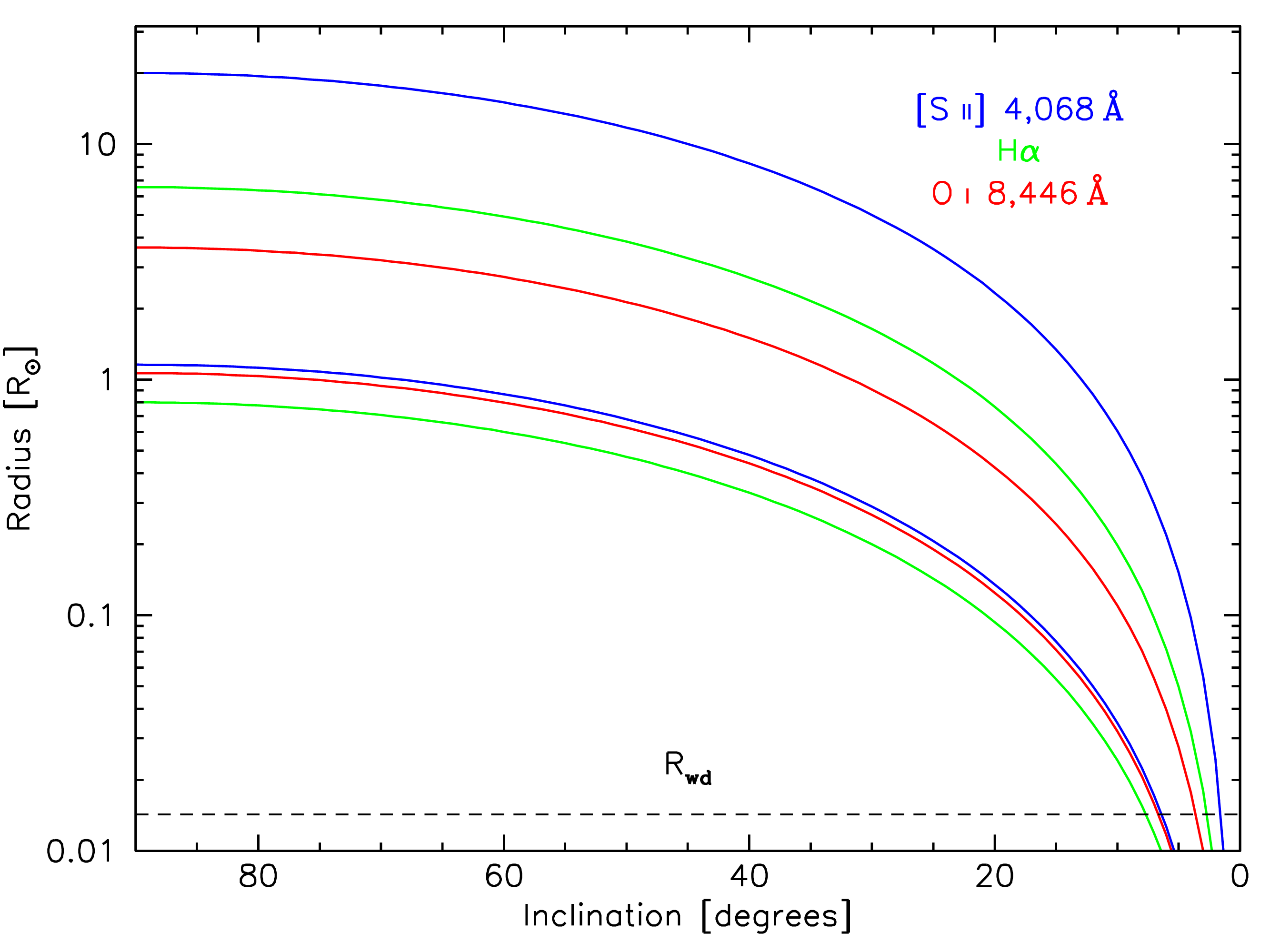}}

\noindent
\textbf{Extended Data Figure\,3.} \textbf{Dynamical constraints on the location of the circumstellar gas emitting the observed double-peaked emission lines.} The gas in the circumstellar disc follows Keplerian orbits, and hence the profile shape of the observed emission lines (see Fig.\,1 and Extended Data Fig.\,2) encodes the location of the gas. The velocity separation of the double-peaks and the maximum velocity in the line wings correspond to motion of gas at the outer edge and inner edge of the disc, respectively. For a given inclination of the disc, these velocities map into semi-major axes. A lower limit on the inclination, $i\ga5^\circ$, arises from the finite size of the white dwarf, and an upper limit on the extent of the disc is provided for an edge-on, $i=90^\circ$, inclination. The forbidden \Line{[S}{ii]}{4,068\,\AA} line has a significantly smaller separation of the double-peaks compared to H$\alpha$ and \Line{O}{i}{8,446\,\AA}, implying a larger radial extent. 

\clearpage
\centerline{\includegraphics[width=120mm]{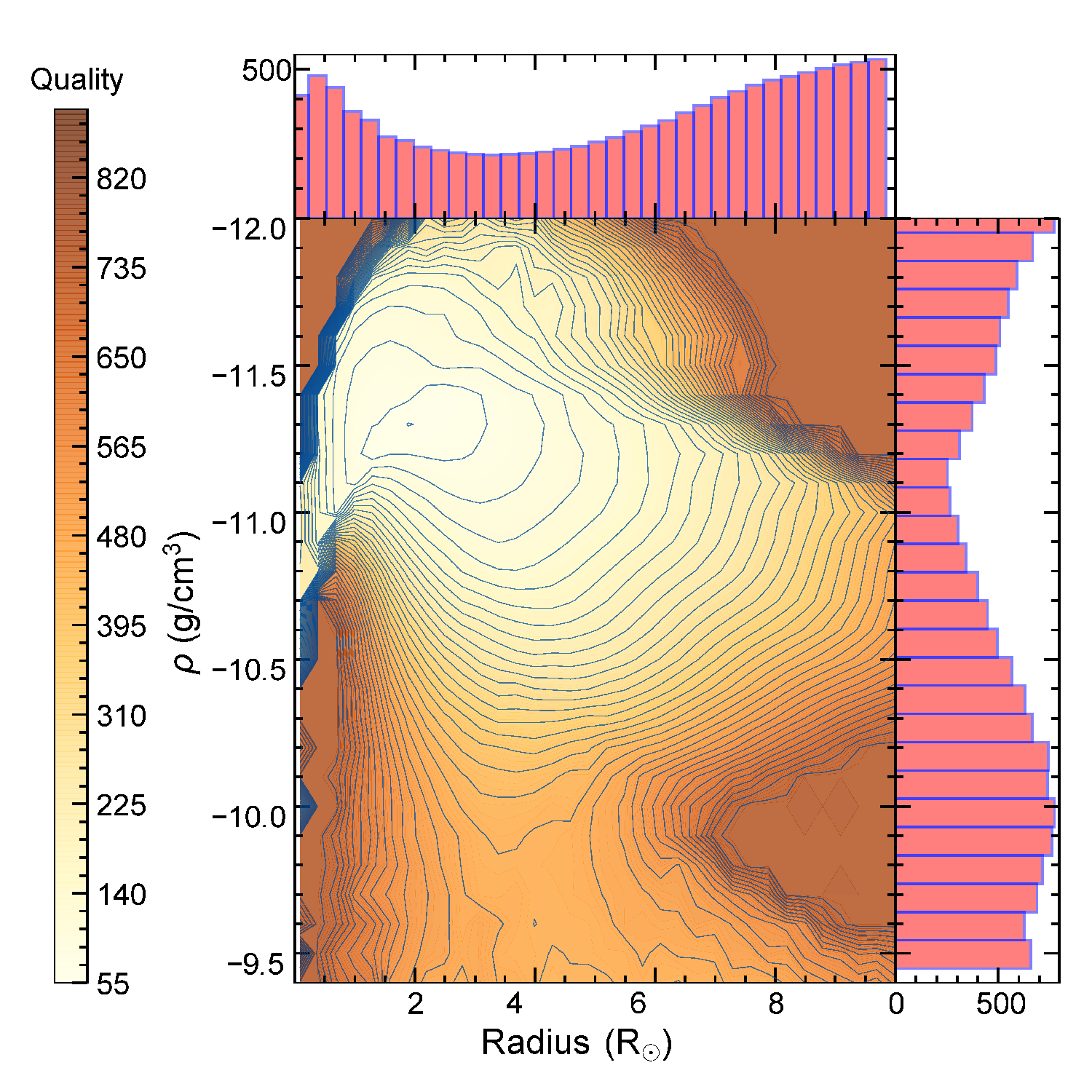}}

\noindent
\textbf{Extended Data Figure\,4.} \textbf{Quality of the \textsc{Cloudy} fits.} The line flux ratios of a grid of \textsc{Cloudy} models spanning a range of gas densities, $\rho$, and distances, $r$, from the white dwarf are compared to the observed values. The two histograms show the average quality for constant $r$ (top) and constant $\rho$ (right). The observed emission line fluxes are reasonably well reproduced by photo-ionised gas with a density $\log(\rho)\simeq-11.3$ and located at $\simeq1-4\,\mathrm{R_\odot}$.

\clearpage
\centerline{\includegraphics[width=120mm]{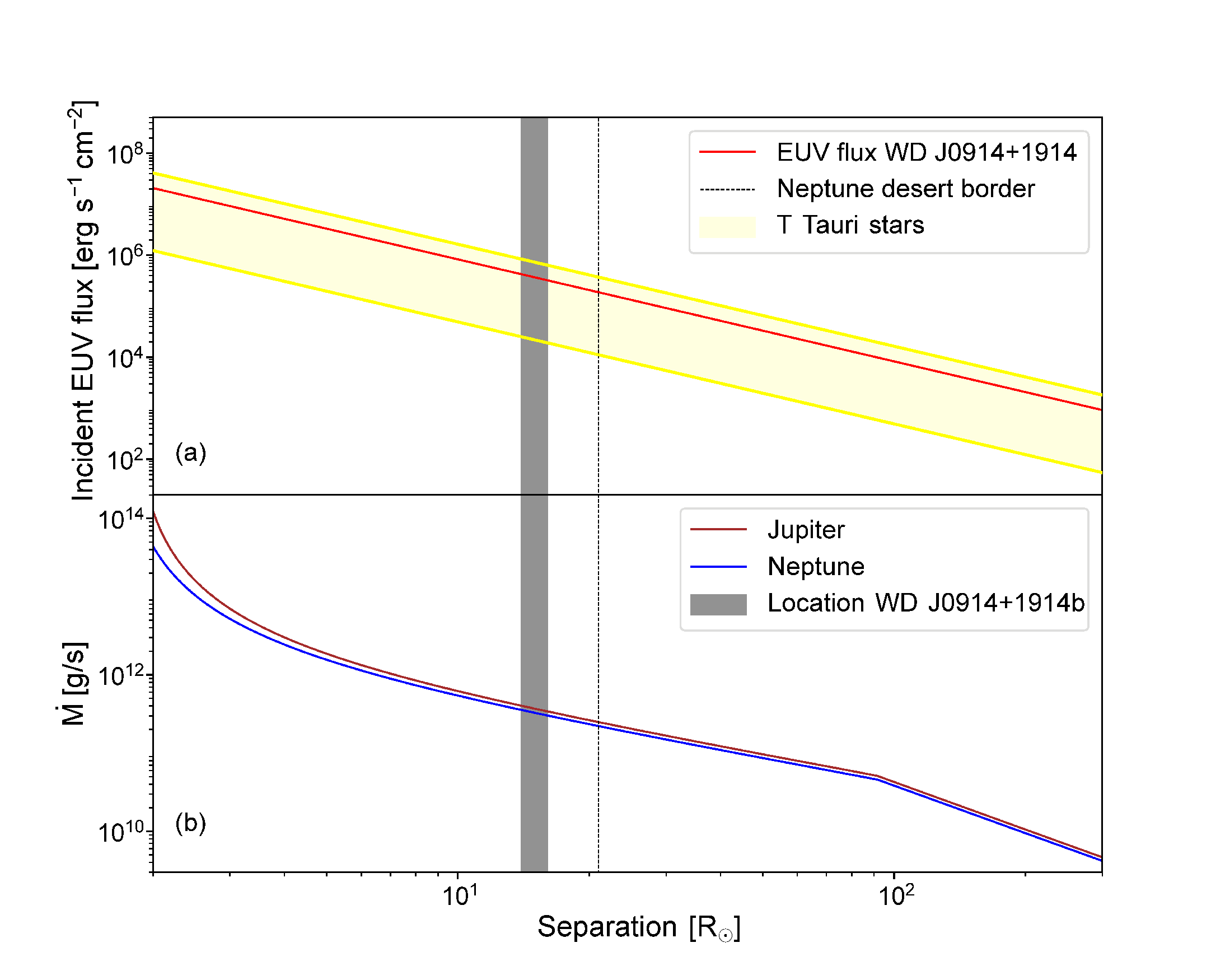}}

\noindent
\textbf{Extended Data Figure\,5.} \textbf{Incident EUV flux (a) and mass loss rates (b) as a function of orbital separation.} (a) Comparison of the irradiating EUV flux around T\,Tauri stars (yellow shaded region) and that of WD\,J0914+1914 (red line). The outer border of the warm Neptune desert is indicated by the vertical dashed line. The orbital separation of the planet orbiting WD\,J0914+1914 estimated from the size of the accretion disc is $\simeq14-16\,\mathrm{R_\odot}$ (grey shaded region). Subject to an EUV luminosity comparable to that of planets around T\,Tauri stars, the giant planet at  WD\,J0914+1914 is well within the warm Neptune desert. (b) Mass loss rates estimated from the assumption of recombination and energy limited hydrodynamic escape for a Jupiter mass and a Neptune mass planet. Significant mass loss could be generated even for separations of up to a few hundred solar radii, well beyond the estimated orbital location of the giant planet at WD\,J0914+1914.  

\clearpage
\centerline{\includegraphics[width=120mm]{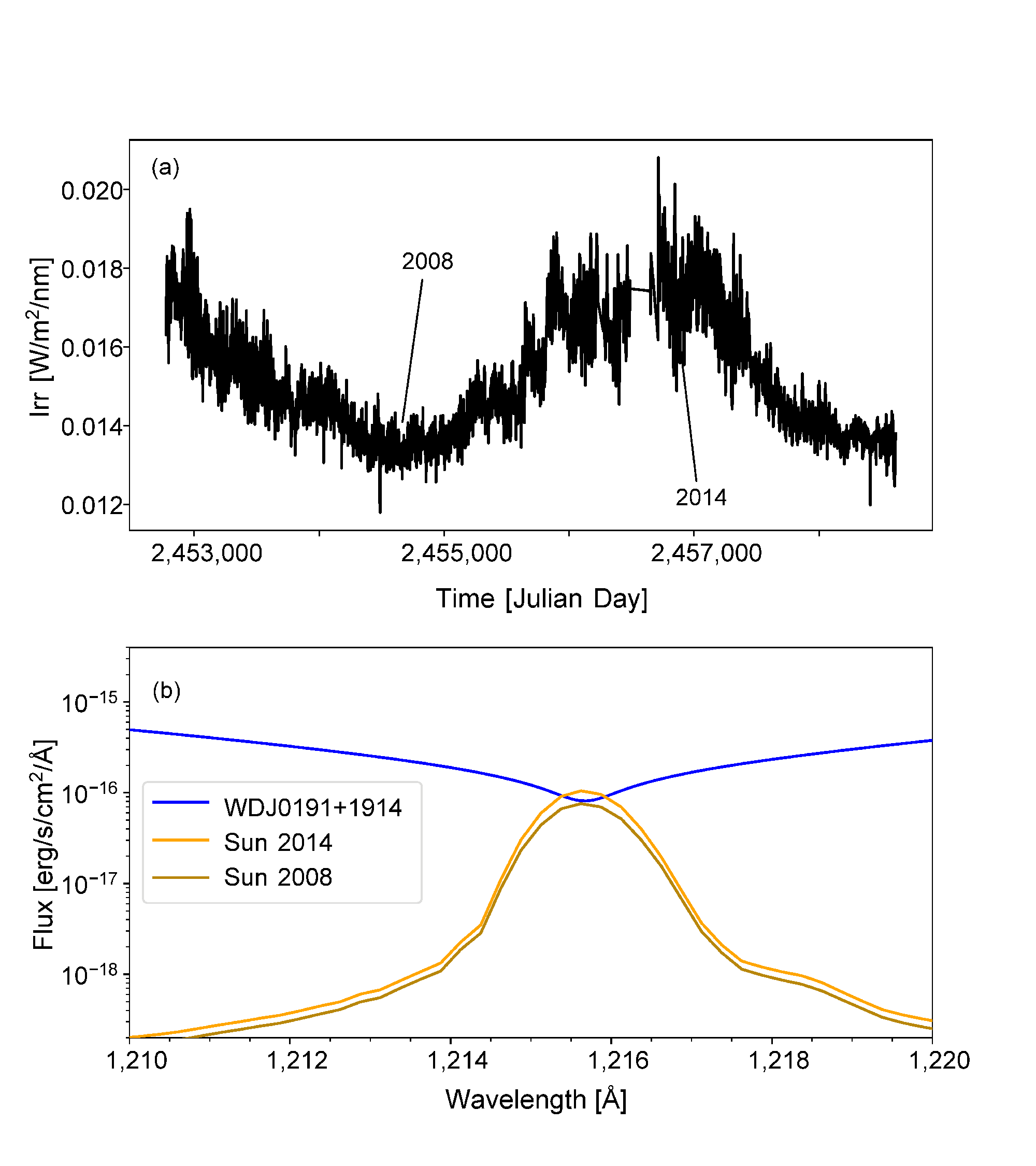}}

\noindent
\textbf{Extended Data Figure\,6.} \textbf{Comparison of the the Ly$\alpha$ emission of WD\,J0914+1914 with the Sun.}
(a) Ly$\alpha$ irradiance of the Sun across a full solar activity cycle as measured by the SORCE SOLSTICE instrument. The radiation pressure on neutral interplanetary hydrogen in the solar system usually exceeds the gravitational force exerted by the Sun. (b) The Ly$\alpha$ flux of the Sun during minimum (2008) and maximum (2014) in comparison to the emission of WD\,J0914+1914 at a distance of $15\,\mathrm{R_\odot}$. Given that WD\,J0914+1914 is less massive than the Sun, and that its  Ly$\alpha$ flux is comparable to that of the Sun in the core of the line, but much larger in the wings (even during the 2014 solar maximum), radiation pressure significantly impedes the inflow of hydrogen, explaining the large depletion of hydrogen with respect to oxygen and sulphur within the circumstellar disc. 

\clearpage
\centerline{\includegraphics[width=120mm]{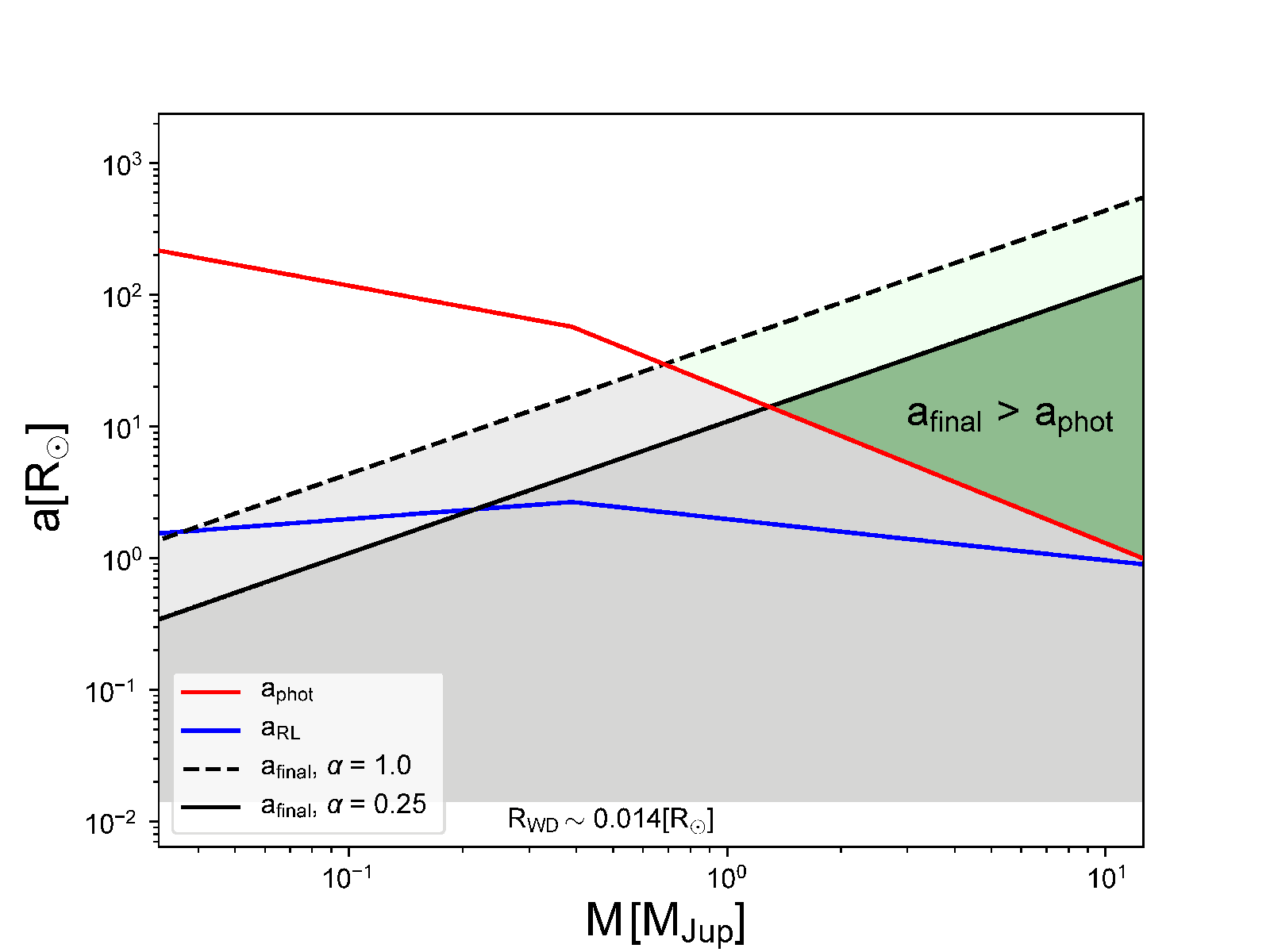}}

\noindent
\textbf{Extended Data Figure\,7.} \textbf{Final separation after common envelope evolution as a function of planetary mass.} We adopted two common envelope efficiencies, $\alpha=0.25$ (solid line), and $\alpha=1.0$ (dashed line). The parameter space of possible outcomes of common envelope evolution lies below these lines (grey shaded region). We consider the smaller efficiency to be more realistic. For configurations below the red line, the planetary mass object will evaporate inside the giant envelope, above the blue line, it would overflow its Roche-lobe. Only planets with parameters within the green shaded region can survive common envelope evolution. Whereas common envelope evolution can bring a Jupiter-mass planet to the estimated location of the planet around WD\,J0914+1914 ($14-16\mathrm{R_\odot}$), smaller planets will be evaporated in the giant envelope. 

\clearpage
\centerline{\includegraphics[width=120mm]{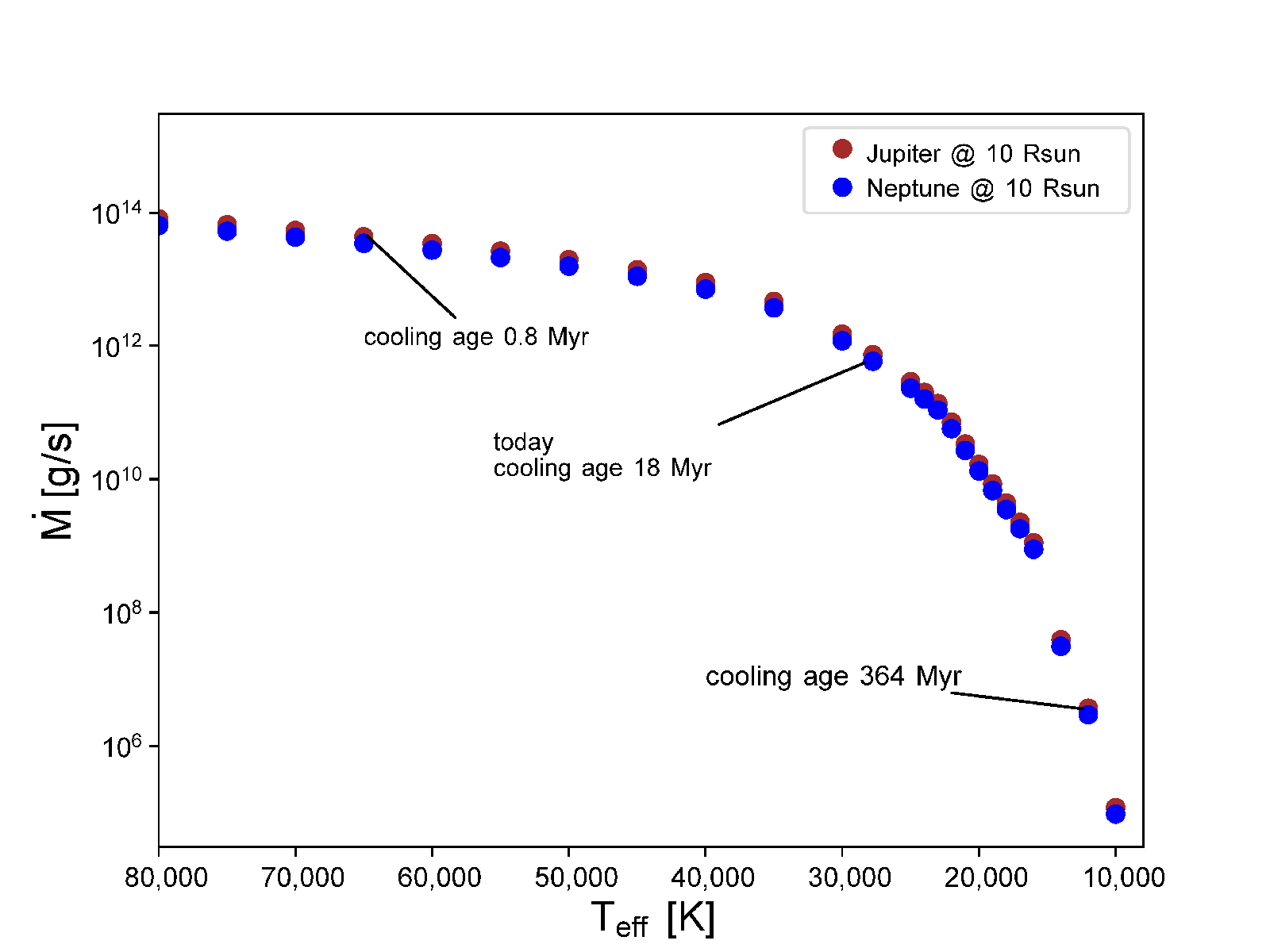}}

\noindent
\textbf{Extended Data Figure\,8.} \textbf{The evolution of the mass loss rate.} White dwarfs cool with time and as a consequence their EUV luminosity decreases. We calculated model spectra for effective temperatures from 80,000\,K to 10,000\,K, integrated the EUV flux, and determined the mass loss rate of a Jupiter and a Neptune at a distance of 10\,R$_{\odot}$. At a cooling age of a $364$\,Myrs the white dwarf will have cooled down to 12,000\,K,  the mass loss rate will drop below $\sim10^6$\,g/s, and the resulting photospheric contamination by oxygen and sulphur will become undetectable. Integrating the mass loss rate over the entire cooling time results in a total mass loss of $\sim0.002$M$_{\mathrm{Jup}}$ which corresponds to $\sim3.7$ per cent of the mass of Neptune.

\clearpage
\spacing{1}
\begin{flushleft}
\textbf{Extended Data Table\,1.} \textbf{White dwarf parameters} 
\begin{tabular}{lc}
\hline
effective temperature $T_\mathrm{eff}$ [K]    & $27743\pm310$ \\
surface gravity $\log g$        [cgs units]   & $7.85\pm0.06$ \\
white dwarf mass $M_\mathrm{wd}$  [$M_\odot$] & $0.56\pm0.03$ \\
cooling age [Myr]                             & $13.3\pm0.5$  \\
progenitor mass [$M_\odot$]                   & $1.0 - 1.6$\\
\textit{Gaia} parallax  [milli-arcsec]        & $2.17\pm0.47$\\
$u_\mathrm{SDSS}$ [mag]                       & $18.629\pm0.026$\\
$g_\mathrm{SDSS}$ [mag]                       & $18.771\pm0.022$\\
$r_\mathrm{SDSS}$ [mag]                       & $19.198\pm0.015$\\
$i_\mathrm{SDSS}$ [mag]                       & $19.529\pm0.022$\\
$z_\mathrm{SDSS}$ [mag]                       & $19.849\pm0.087$\\
SDSS spectroscopic identifiers [MJD-PLT-FIBER] & $53700-2286-0021$ \\
                                              & $56017-5768-0660$\\
\hline
\end{tabular}
\end{flushleft}

\clearpage
\begin{flushleft}
\textbf{Extended Data Table\,2.} \textbf{Element number abundances (log[Z/H])} 
\spacing{1}
\begin{tabular}{lrrrr}
\hline
Element             & photosphere & disk & disk   & solar \\
                    &             &      & scaled & \\
\hline
He & $<-2.1$        &             &          & $-1.07$ \\
C  & $<-4.8$        & $<-1.17$    & $<-4.71$ & $-3.57$ \\
N  & $<-3.7$        & $<-1.00$    & $<-4.54$ & $-4.17$ \\
O  & $-3.25\pm0.2$  & $0.29\pm0.3$& $-3.25$  & $-3.31$ \\
Na & $<-4.3$        & $<-3.85$    & $<-7.39$ & $-5.76$ \\
Mg & $<-5.8$        & $<-2.10$    & $<-5.64$ & $-4.40$ \\
Al & $<-6.0$        & $<-2.70$    & $<-6.24$ & $-5.55$ \\
Si & $<-5.2$        & $<-3.19$    & $<-6.73$ & $-4.49$ \\
S  & $-4.15\pm0.2$ & $-0.21\pm0.3$ & $-3.75$ & $-4.88$ \\
K  & $<-5.2$        & $<-3.67$    & $<-7.21$ & $-6.97$ \\
Ca & $<-6.0$        & $<-6.18$    & $<-9.72$ & $-5.66$ \\
Mn & $<-4.4$        &             &         & $-6.57$ \\
Fe & $<-4.2$        & $<-4.03$    & $<-7.57$ & $-4.50$ \\
Zn & $<-3.0$        &             &         & $-7.44$ \\[0.1ex]
\hline
\end{tabular}

\end{flushleft}

\medskip
\noindent
\spacing{2}
The number abundances in the white dwarf photosphere were derived from fitting model spectra to the oxygen and sulphur lines detected in the X-Shooter spectrum. Upper limits were obtained from the non-detection of the strongest lines of the individual elements. The number abundances in the circumstellar disc were derived from fitting \textsc{Cloudy} models to the observed flux ratios of the emission lines of hydrogen, oxygen, and sulphur, and upper limits for the remaining elements were obtained from the non-detection of corresponding emission lines. Hydrogen is significantly depleted in the disc. To facilitate the comparison between these two independent measurements, the column ``disc scaled'' gives the abundances and upper limits obtained from the model of the gaseous disc scaled to match the photospheric oxygen abundance. Solar number abundances are provided as reference.

\end{document}